\documentclass[onecolumn,12pt]{IEEEtran}

\usepackage{cite}

\ifCLASSINFOpdf
\usepackage[pdftex]{graphicx}
\else

\usepackage[dvips]{graphicx}
\DeclareGraphicsExtensions{.eps}
\fi
\usepackage[cmex10]{amsmath}
\usepackage{amssymb}
\usepackage{amsmath}
\usepackage{amsthm}
\usepackage{mathrsfs}
\usepackage{graphicx}
\usepackage{algorithm}
\usepackage{algpseudocode}
\usepackage{subfig}
\usepackage{bibspacing}
\usepackage{cite}
\usepackage{bbm}
\usepackage{geometry}
\newtheorem{theorem}{Theorem}
\newtheorem{remark}{Remark}

\newtheorem{lemma}{Lemma}
\newtheorem{proposition}{Proposition}

\newcommand{\Rmnum}[1]{\expandafter\@slowromancap\romannumeral #1@}

\ifCLASSINFOpdf

\else

\fi

\begin{document}

\title{Polar Codes for Broadcast Channels with Receiver Message Side Information and Noncausal State Available at the Encoder}
\author{\IEEEauthorblockN{Jin~Sima 
and Wei~Chen}
}

\maketitle
\vspace{1cm}

\maketitle
\thispagestyle{empty}
\pagestyle{empty}
\begin{abstract}
In this paper polar codes are proposed for two receiver broadcast channels with receiver message side information (BCSI) and noncausal state available at the encoder, referred to as BCSI with noncausal state for short, where the two receivers know a priori the private messages intended for each other. This channel generalizes BCSI with common message and Gelfand-Pinsker problem and has applications in cellular communication systems. We establish an achievable rate region for BCSI with noncausal state and show that it is strictly larger than the straightforward extension of the Gelfand-Pinsker result. To achieve the established rate region with polar coding, we present polar codes for the general Gelfand-Pinsker problem, which adopts chaining construction and utilizes causal information to pre-transmit the frozen bits. It is also shown that causal information is necessary to pre-transmit the frozen bits. Based on the result of Gelfand-Pinsker problem, we use the chaining construction method to design polar codes for BCSI with noncausal state. The difficulty is that there are multiple chains sharing common information bit indices. To avoid value assignment conflicts, a nontrivial polarization alignment scheme is presented. It is shown that the proposed rate region is tight for degraded BCSI with noncausal state.
\end{abstract}


\IEEEpeerreviewmaketitle
\section{Introduction}
In Arikan's pioneering work \cite{arikan2009channel}, he introduced polar codes, which constitute a new and promising class of practical capacity achieving codes. By exploiting the channel/source polarization phenomenon, polar codes are capable of achieving channel capacity with encoding and decoding complexity $O(n\log n)$ and error probability $O(2^{-n^{\beta}})$ \cite{arikan2009channel,arikan2009rate}. Polar codes, which are originally proposed for symmetric binary-input memoryless channels, have been richly investigated and generalized to various channel/source coding problems. The works in \cite{sasoglu2009polarization,mori2010channel} extended polar codes for arbitrary finite input alphabet size. Polar codes for asymmetric channels were proposed in \cite{honda2013polar,sutter2012achieving}, and in \cite{goela2013polar} in the treatment on broadcast channels. For multi-user scenarios, polar codes were studied for multiple access channels \cite{abbe2012polar,sasoglu2013polar,mahdavifar2013achieving}, broadcast channels \cite{goela2013polar,mondelli2014achieving,andersson2013polar}, interference channels \cite{6874845,appaiah2011polar}, wiretap channels \cite{mahdavifar2011achieving,koyluoglu2012polar,sasoglu2013new}, relay channels \cite{blasco2012polar}, Gelfand-Pinsker problem \cite{korada2009polar,gad2014asymmetric,burshtein}, and lossless and lossy source coding problems \cite{korada2009polar,arikan2010source,korada2010polar}.

In the work \cite{goela2013polar}, Goela, Abbe, and Gastpar introduced polar codes for realizing superposition strategy and Marton's strategy, which comprise the main coding strategies for broadcast channels. To guarantee the alignment of polarization indices, the coding scheme requires some degradedness conditions with respect to the auxiliary random variables and channel outputs. Such degradedness requirements can be removed by adopting the polarization alignment techniques proposed by Mondelli, Hassani, Sason, and Urbanke \cite{mondelli2014achieving}, where multi-block transmission and block chaining are considered. The work in \cite{andersson2013polar} proposed polar codes for two receiver broadcast channels with receiver message side information (BCSI), where each receiver knows the message intended for the other. The BCSI naturally arises in two-way communication in cellular systems, where a pair of users exchange messages with each other through the help of the base station.
Two way communication consists of the multiple access uplink transmission and the broadcasting downlink transmission. Since the pair of users that exchange messages with each other know side information about their own messages, the downlink transmission to them can be modeled as BCSI. It is found that polar coding combined with network coding is able to utilize the receiver side information and achieve the capacity regions for the symmetric BCSI and symmetric BCSI with common and confidential messages \cite{andersson2013polar}.

In this paper, we consider polar codes for BCSI with common message and with noncausal state available at the encoder, which is a generalization of Gelfand-Pinsker channel and BCSI. The motivation for the study of such channel is that the channel arises in multi-user cellular communication systems with two-way communication tasks or pairwise message exchange requests. For each pair of users that exchange messages, broadcasting to them in the downlink transmission can be regarded as BCSI with noncausal state, by considering the interference from signals of other users as noncausal state known at the base station. The application of coding for BCSI with noncausal state were proposed in \cite{oechtering2010achievable,7037031} to tackle the interference that presents in multi-user cellular communication systems. BCSI with noncausal state was studied in a previous work \cite{6294445}, where a coding scheme combining Gelfand-Pinsker binning and network coding was proposed. Its related scenarios, broadcast channels with noncausal state, has received much attention and has been investigated in, e.g., \cite{steinberg2005coding,nair2010achievability,6089375}.

Polar codes for Gelfand-Pinsker problems have been presented. Polar codes for binary channels with additive noise and interference was proposed in \cite{korada2009polar}. Noisy write once memory was considered in \cite{gad2014asymmetric}, where polar codes with polynomial computational and storage complexity were proposed. For general Gelfand-Pinsker settings, the work in \cite{gad2014asymmetric,burshtein} proposed polar coding schemes based on the the block chaining method in \cite{mondelli2014achieving}. The problem of applying the chaining construction to the Gelfand-Pinsker settings is to communicate the state information to the receiver in the first block. This problem was not addressed in \cite{burshtein}. The work in \cite{gad2014asymmetric} proposed a solution to this problem by using an extra phase to transmit the frozen bits in the first block, where the channel state information is not used by the encoder. As we will show in the next, this solution may not work in some cases. In particular, the state information is needed by the encoder to transmit the frozen bits in the first block.

In this paper, we establish an achievable rate region for BCSI with common message and with noncausal state. Polar coding schemes are presented to achieve the established region. To achieve this, we first propose polar codes for the general Gelfand-Pinsker problem, based on the block chaining construction in \cite{mondelli2014achieving}. A pre-communication phase that utilizes causal state information is performed to transmit the frozen bits in the first block. It is also shown that the state information is necessary to transmit these frozen bits. We then use the result in the Gelfand-Pinsker problem to construct polar codes for BCSI with noncausal state. The chaining construction is employed with nontrivial polarization alignment since there are two chains sharing common information bit indices in order to perform Gelfand-Pinsker coding simultaneously for the two users.
To overcome the problem that the two chains may overlap and cause value assignment conflicts,
the two chains are generated in opposite directions so that the overlapped sets only needs to carry the XOR of the bits contained in the two chains.
We present an example to show that it is strictly larger than the existing achievable rate region \cite{6294445}. It is shown that the established rate region is tight for degraded BCSI with common message and with noncausal state.

The proposed polar coding schemes have the same performance as polar codes for point to point channels, that is, encoding and decoding complexity $O(n\log n)$ and error probability $O(2^{-n^\beta})$ for $0<\beta<\frac{1}{2}$. In this paper we consider binary inputs for channels. The extension to higher input alphabet size can be similarly made following the techniques in \cite{sasoglu2009polarization,mori2010channel}.

The rest of the paper is organized as follows. In section \uppercase\expandafter{\romannumeral2} channel model and some notations are presented. For polar coding schemes, we begin with polar codes for BCSI with common message in section \uppercase\expandafter{\romannumeral3}. In section \uppercase\expandafter{\romannumeral4} we propose polar codes for the general Gelfand-Pinsker settings and use the result to construct a polar coding scheme for BCSI with noncausal state. Section \uppercase\expandafter{\romannumeral5} presents summaries of this paper.
\section{Models and Notations}

\subsection{Channel Model}
\begin{figure}\label{fig:1}
\center
\includegraphics[scale=0.7]{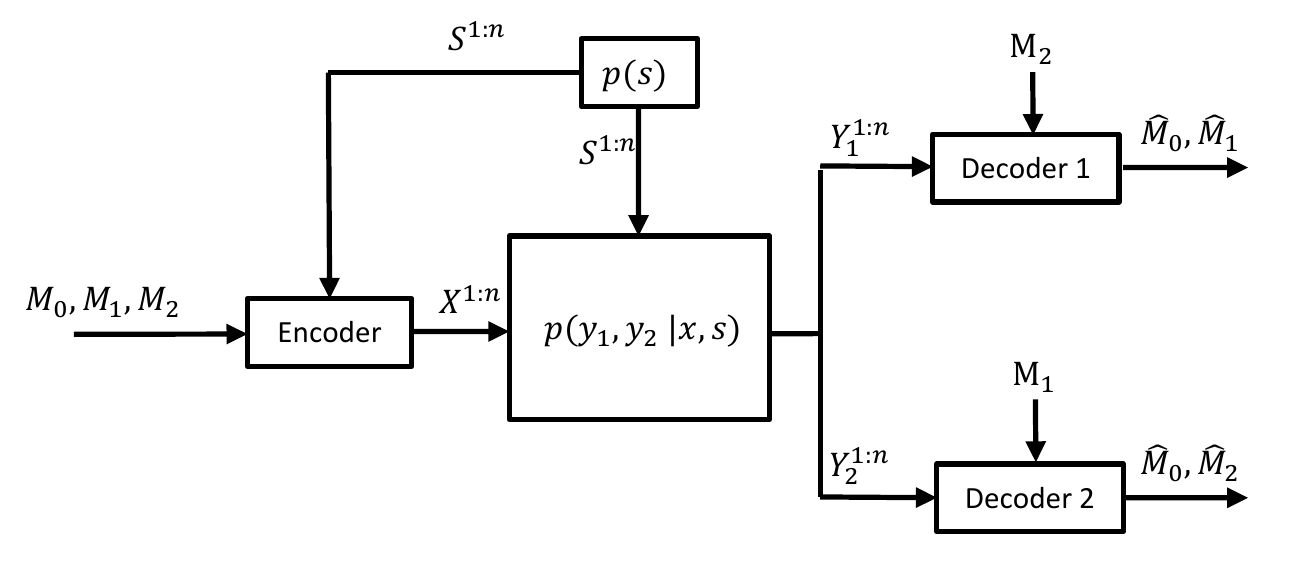} 
\caption{BCSI with noncausal state}
\setlength\belowcaptionskip{0pt}
\end{figure}
Broadcast channels with receiver message side information (BCSI) and with noncausal state available at the encoder (as shown in Fig. $1$), which is referred to as BCSI with noncausal state for short, is a two-receiver discrete memoryless broadcast channels (DMBC) with state
\begin{equation}\label{channelmodel}
(\mathcal{X}\times\mathcal{S},P_{Y_1,Y_2|X,S}(y_1,y_2|x,s),\mathcal{Y}_1\times\mathcal{Y}_2),
\end{equation}
with input alphabet $\mathcal{X}$, state alphabet $\mathcal{S}$, output alphabets $\mathcal{Y}_1,\mathcal{Y}_2$ and conditional distribution $P_{Y_1,Y_2|X,S}$ $(y_1,y_2|x,s)$. The channel state sequence $S^{1:n}$ is a sequence of $n$ i.i.d. random variables with pmf $P_{S}(s)$ and is noncausally available at the encoder. The sender wishes to send a message tuple $(M_0,M_1,$ $M_2)\in[1:2^{nR_0}]\times[1:2^{nR_1}]\times [1:2^{nR_2}]$ to receivers $1$ and $2$, where receivers $1$ and $2$ know side information of messages $M_2$ and $M_1$ respectively. $M_0$ is a common message intended for both receivers.

A $(2^{nR_0},2^{nR_1},2^{nR_2},n)$ code consists of a message set $[1:2^{nR_0}]\times[1:2^{nR_1}]\times [1:2^{nR_2}]$, an encoder
$\zeta : [1:2^{nR_0}]\times[1:2^{nR_1}]\times [1:2^{nR_2}]\times \mathcal{S}^n \rightarrow \mathcal{X}^{n}$ that maps $(M_0,M_1,M_2,S^{1:n})$ to a codeword $X^{1:n}$, and
two decoders $\xi_1:\mathcal{Y}^{n}_1\times [1:2^{nR_2}]\rightarrow [1:2^{nR_0}]\times[1:2^{nR_1}]$ and $\xi_2:\mathcal{Y}^{n}_2\times [1:2^{nR_1}]\rightarrow [1:2^{nR_0}]\times[1:2^{nR_2}]$ that map $(Y^{1:n}_1,M_2)$ and $(Y^{1:n}_2,M_1)$ to $(\hat{M}_0,\hat{M}_1)$ and $(\hat{M}_0,\hat{M}_2)$ respectively. Here $Y^{1:n}_i$ is the received sequence of receiver $i$.
A rate tuple $(R_0,R_1,R_2)$ is achievable if there exists a $(2^{nR_0},2^{nR_1},2^{nR_2},n)$ code such that the average error probability of the code
\begin{equation}
P^{(n)}_e=P\{\xi_1(Y^{1:n}_1,M_{2})\ne \{M_0,M_1\} \cup \xi_2(Y^{1:n}_2,M_1)\ne \{M_0,M_2\}\}
\end{equation}
tends to zero as $n$ goes to infinity. The capacity region $\mathcal{C}$ is the closure of the set of all achievable rate tuples $(R_0,R_1,R_2)$.

For each random variable $U$, we shall use the notation $U^{1:n}$ to denote the sequence of $n$ i.i.d. random variables drawn from pmf $P_{U}(u)$. The $i$-th element of $U^{1:n}$ is denoted as $U^i$.
\subsection{Polarization}
Let $(X,Y)\sim P_{X,Y}$ be a pair of random variables with alphabet $\mathcal{X}\times\mathcal{Y}$, where $\mathcal{X}=\{0,1\}$ and $\mathcal{Y}$ is an arbitrary finite set.
The Bhattacharyya parameter $Z(X|Y)\in[0,1]$ with respect to $(X,Y)$ is defined as
\begin{equation}
Z(X|Y)=2\sum_{y\in\mathcal{Y}}P_{Y}(y)\sqrt{P_{X|Y}(0|y)P_{X|Y}(1|y)}
\end{equation}
The following lemma establishes upper and lower bounds of the conditional entropy $H(X|Y)$ in terms of the Bhattacharyya parameter $Z(X|Y)$.
\begin{proposition}
\textup{[20, Proposition 2] For a pair of random variables $(X,Y)\sim P_{X,Y}$, where $X\in\{0,1\}$, and $Y$ takes values in a finite alphabet, we have
\begin{equation}
\begin{split}
&Z(X|Y)^2\le H(X|Y),~~~~H(X|Y)\le \log_2(1+Z(X|Y))
\end{split}
\end{equation}
}
\end{proposition}
For $n=2^k$, $(X^{1,n},Y^{1:n})=\Big ((X^1,Y^1),\ldots,(X^n,Y^n)\Big )$ is a sequence of $n$ i.i.d. copies of random variables $(X,Y)$.  Let the sequence $U^{1:n}$ be $U^{1:n}=X^{1:n}G_n$, where $G_n={\left( {\begin{array}{*{20}{c}}
1&0\\
1&1
\end{array}} \right)^{ \otimes k}}$ is the polar matrix and $\otimes$ denotes the Kronecker power.
\begin{proposition}
\textup{
For a constant $\beta$ that satisfies $0<\beta <\frac{1}{2}$,
\begin{equation}
\begin{split}
&\lim_{n\rightarrow \infty}\frac{1}{n}|\{i\in[n]:Z(U^i|Y^{1:n},U^{1:i-1})\ge 1-2^{-n^\beta}\}|=H(X|Y),\\
&\lim_{n\rightarrow \infty}\frac{1}{n}|\{i\in[n]:Z(U^i|Y^{1:n},U^{1:i-1})\le 2^{-n^\beta}\}|=1-H(X|Y).
\end{split}
\end{equation}
Specially, when $Y$ is constant, we have
\begin{equation}
\begin{split}
&\lim_{n\rightarrow \infty}\frac{1}{n}|\{i\in[n]:Z(U^i|U^{1:i-1})\ge 1-2^{-n^\beta}\}|=H(X),\\
&\lim_{n\rightarrow \infty}\frac{1}{n}|\{i\in[n]:Z(U^i|U^{1:i-1})\le 2^{-n^\beta}\}|=1-H(X).
\end{split}
\end{equation}
}
\end{proposition}
The proof of this proposition is given in [5, Theorem 1]. The proposition can also be proved by defining a super-martingale with respect to the Bhattacharyya parameter, as mentioned in \cite{goela2013polar}.

Based on the above polarization phenomenon, which implies that the channel $W^i=W(U^i|Y^{1:n},$ $U^{1:i-1})$ either becomes deterministic or becomes rather noisy, polar codes can be designed to achieve channel capacity with low complexity and low error probability. For an information set $\mathcal{I}$, the encoder puts message information in the bits $u^{\mathcal{I}}=(u^{i}: i\in\mathcal{I})$, and generates the frozen bits $u^{\mathcal{I}^c}=(u^{i}: i\in\mathcal{I}^c)$ according to a set of randomly chosen maps $\lambda(u^{1:i-1})$ where the randomness is shared between the encoder and the decoders. Note that shared randomness is not necessary in generating the frozen bits, as pointed out in \cite{gad2014asymmetric}, where polar coding schemes that avoid using large boolean functions are proposed. After generating the sequence $U^{1:n}$, the encoder transmits $U^{1:n}G^{-1}_n=U^{1:n}G_n$ as the channel input. The decoder adopts successive decoding to recover the sequence $u^{1:n}$. It is shown that the probability of error decays like $O(2^{-n^\beta})$ for $0<\beta<\frac{1}{2}$ and the encoding/decoding complexity is $O(n\log n)$.
\section{Polar Codes for BCSI with Common Message}
To demonstrate our polar code scheme for BCSI with noncausal state, we begin in this section with a simpler case of broadcast channels with receiver message side information (BCSI) and with common message, which can be viewed as BCSI with common message and with constant state.
It has been proved the capacity region for BCSI with common message is given by \cite{kramer2007capacity}
\begin{equation}\label{crforbbc}
\begin{split}
R_1+R_0\le I(X;Y_1),
~~~R_2+R_0\le I(X;Y_2).
\end{split}
\end{equation}
The following theorem shows the achievability of the rate region \eqref{crforbbc} by using polar codes.
\begin{theorem}
\textup{Consider a BCSI $(\mathcal{X},P_{Y_1,Y_2|X}(y_1,y_2|x),\mathcal{Y}_1\times\mathcal{Y}_2)$ with binary input alphabet $\mathcal{X}=\{0,1\}$, for any rate tuple $(R_0,R_1,R_2)$ satisfying \eqref{crforbbc},
there exists a polar code sequence with block length $n$ that achieves $(R_0,R_1,R_2)$. As $n$ increases, the encoding and decoding complexity is $O(n\log n)$ and the error probability is $O(2^{-n^\beta})$ for any $0<\beta<\frac{1}{2}$.
}
\end{theorem}
In the rest of this section, we deal with the proof of theorem $1$, namely, the coding scheme and the complexity and error analyses.
Let $X^{1:n}$ be a sequence of $n$ i.i.d. variables with pmf $P_{X}(x)$. Set the sequence $U^{1:n}=X^{1:n}G_n$. Define the polarization sets
\begin{equation}
\begin{split}
&\mathcal{H}^{(n)}_{U}=\{i\in[n]:Z(U^i|U^{1:i-1})\ge 1-2^{-n^\beta}\},\\
&\mathcal{L}^{(n)}_{U}=\{i\in[n]:Z(U^i|U^{1:i-1})\le 2^{-n^\beta}\},\\
&\mathcal{H}^{(n)}_{U|Y_1}=\{i\in[n]:Z(U^i|Y_1^{1:n},U^{1:i-1})\ge 1-2^{-n^\beta}\},\\
&\mathcal{L}^{(n)}_{U|Y_1}=\{i\in[n]:Z(U^i|Y_1^{1:n},U^{1:i-1})\le 2^{-n^\beta}\},\\
&\mathcal{H}^{(n)}_{U|Y_2}=\{i\in[n]:Z(U^i|Y_2^{1:n},U^{1:i-1})\ge 1-2^{-n^\beta}\},\\
&\mathcal{L}^{(n)}_{U|Y_2}=\{i\in[n]:Z(U^i|Y_2^{1:n},U^{1:i-1})\le 2^{-n^\beta}\}.
\end{split}
\end{equation}
Let the information sets for users $1$ and $2$ be
\begin{equation}\label{informationsetbc}
\begin{split}
&\mathcal{I}_1=\mathcal{H}^{(n)}_{U}\cap\mathcal{L}^{(n)}_{U|Y_1},~~~~\mathcal{I}_2=\mathcal{H}^{(n)}_{U}\cap\mathcal{L}^{(n)}_{U|Y_2},\\
\end{split}
\end{equation}
which indicates that the bit $U^{i}$ with $i\in\mathcal{I}_m,~m=1,2$ is distributed almost uniformly and independently of $U^{1:i-1}$ and can be deduced by using the received sequence $Y^{1:n}_m$ and sequence $U^{1:i-1}$. Note that $\mathcal{H}^{(n)}_{U|Y_1}\subseteq \mathcal{H}^{(n)}_{U}$ and $|\mathcal{H}^{(n)}_{U|Y_1}\cup\mathcal{L}^{(n)}_{U|Y_1}|=n-o(n)$.
According to Proposition $2$, the following result holds.
\begin{proposition}
\textup{For the information sets $\mathcal{I}_1$ and $\mathcal{I}_2$, we have
\begin{equation}
\begin{split}
&\lim_{n\rightarrow \infty}\frac{|\mathcal{I}_1|}{n}=I(X;Y_1),~~~~\lim_{n\rightarrow \infty}\frac{|\mathcal{I}_2|}{n}=I(X;Y_2),
\end{split}
\end{equation}
}
\end{proposition}

\subsection{Polar Coding Protocol}
Similar to polar codes for point-to-point channels, the encoder puts the information of $(M_0,M_1)$ and $(M_0,M_2)$ into bits $u^{\mathcal{I}_1}$ and $u^{\mathcal{I}_2}$ respectively. The bits $u^{(\mathcal{I}_1\cap\mathcal{I}_2)^c}$ are frozen and generated by using randomized maps, where the randomness is shared between the encoder and the decoders so that each user $m=1,2$ can decode out the full sequence $u^{1:n}$ once $u^{\mathcal{I}_1\cup\mathcal{I}_2}$ is determined.

For the case when $R_0=0$, the above strategy can be done with the help of network coding \cite{andersson2013polar}.
The encoder puts the bitwise XOR of $M_1$ and $M_2$ message bits in $u^{\mathcal{I}_1\cap\mathcal{I}_2}$. Since users $1$ and $2$ know the messages intended for each other, both users can recover the bits $u^{\mathcal{I}_1\cup\mathcal{I}_2}$ and hence the sequence $u^{1:n}$.
When $nR_0>|\mathcal{I}_1\cap\mathcal{I}_2|$ (this may happen when, say, $R_0\ne 0$ and $\mathcal{I}_1\cap\mathcal{I}_2=\emptyset$), part of the $M_0$ message bits has to be transmitted via the bits $u^{\mathcal{I}_1-\mathcal{I}_2}$ and $u^{\mathcal{I}_2-\mathcal{I}_1}$. In this case receiver $m,~m=1,2$ may not decode the bits $u^{\mathcal{I}_{3-m}-\mathcal{I}_{m}}$ since neither it knows the message $M_0$ nor can it recover the bits $u^{\mathcal{I}_{3-m}-\mathcal{I}_{m}}$ correctly with its received sequence $y^{1:n}_m$. To deal with such cases, we adopt the block chaining construction presented in \cite{mondelli2014achieving}.

Without loss of generality it is assumed that $R_1\ge R_2$. Split the message $M_1$ into $M_{11}$ and $M_{10}$ at rates $R_{11}$ and $R_{10}$ respectively such that $R_{10}=R_2$.
Let $M_0'=(M_0,M_{10}\oplus M_{2})$ be a new equivalent common message, where $\oplus$ denotes the bitwise XOR operation.
Note that user $1$ and $2$ can recover their desired messages by decoding $(M_0',M_{11})$ and $M_0'$ respectively. The message rates $(R_0,R_1,R_2)$ satisfy $R_1=R_0'+R_{11}$, $R_2+R_0=R_0'$. Define the sets
\begin{equation}
\begin{split}
&\mathcal{D}_1=\mathcal{I}_1-\mathcal{I}_2,~~~~\mathcal{D}_2=\mathcal{I}_2-\mathcal{I}_1.\\
\end{split}
\end{equation}
Let $\mathcal{D}_{10}$ be a subset of $\mathcal{D}_1$ such that $|\mathcal{D}_{10}|=|\mathcal{D}_2|$. The coding scheme consists of $k$ blocks. In block $1$, bits $u^{\mathcal{I}_2}$ are inserted with the $M'_0$ information and bits $u^{\mathcal{D}_1}$ are generated by using randomized maps with randomness shared between the encoder and the decoders. For block $j=2,\ldots,k$, the encoder puts the $M_{11}$ information in bits $u^{\mathcal{D}_{1}\backslash\mathcal{D}_{10}}$ and fills the bits $u^{\mathcal{D}_{10}}$ with the information contained in $u^{\mathcal{D}_2}$ in block $j-1$. In block $j=2,\ldots,k-1$, the bits $u^{\mathcal{I}_2}$ are filled with $M'_0$ message bits. In block $k$, the encoder puts $M'_0$ information in the bits $u^{\mathcal{I}_1\cap\mathcal{I}_2}$ and generates the
bits $u^{\mathcal{D}_2}$ according to randomized maps.
The scheme is presented in Fig. $2$.
\begin{figure}\label{fig:1}
\center
\includegraphics[scale=0.7]{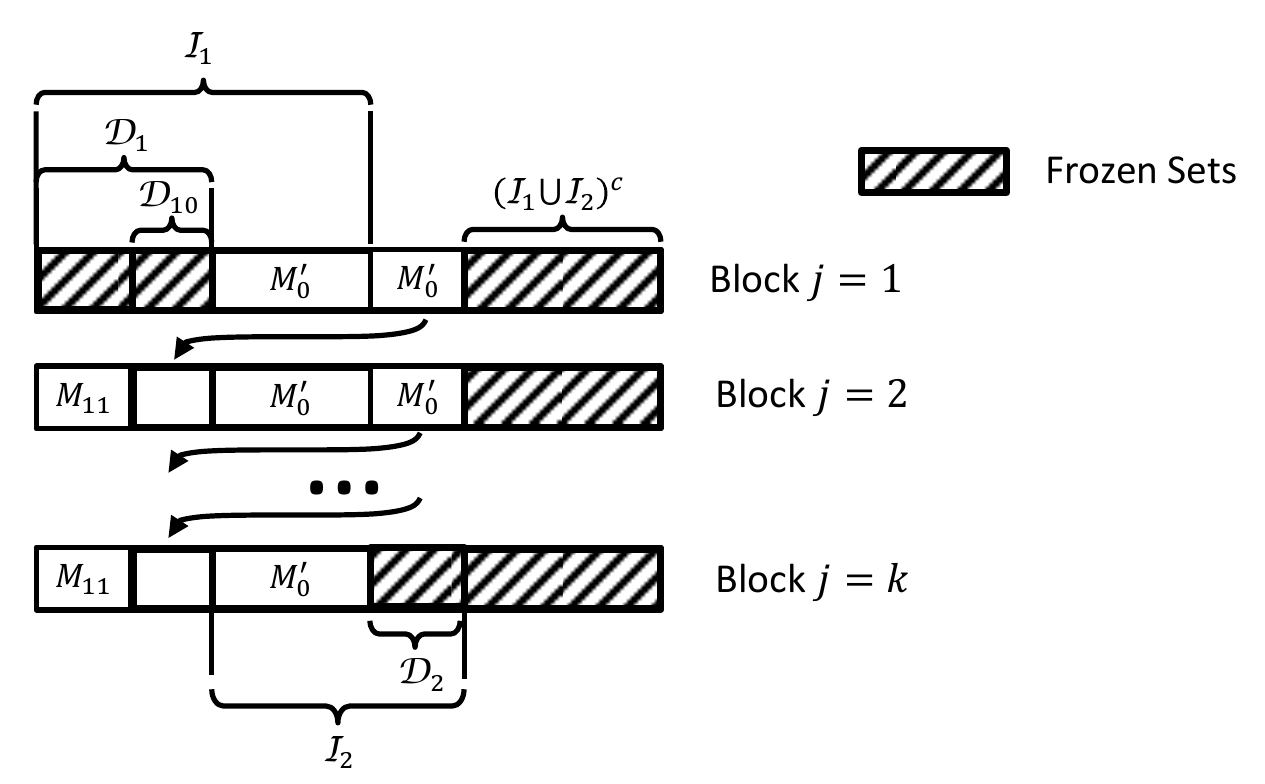} 
\caption{Polar coding scheme for BCSI with common message}
\setlength\belowcaptionskip{0pt}
\end{figure}

Upon decoding, user $2$ starts from block $1$ to block $k$. As user $2$ decodes, the bits $u^{\mathcal{D}_{10}}$ can be recovered since the content therein is contained in the bits $u^{\mathcal{D}_2}$ decoded in the last block (The bits $u^{\mathcal{D}_{10}}$ in block $1$ can be decided by using the pre-determined randomized map). Meanwhile, the bits $u^{\mathcal{D}_{1}-\mathcal{D}_{10}}$ are available at user $2$ since they are filled with $M_{11}$ messages. The bits $u^{\mathcal{I}_2}$ can be decoded based on the received sequence $y^{1:n}_2$. The remaining bits $u^{(\mathcal{I}_1\cup\mathcal{I}_2)^c}$ can be calculated using the shared randomized maps. Therefore, user $2$ can decode $u^{1:n}$ successfully. Similarly, user $1$ starts from block $k$ to block $1$ and is able to decode the sequence $u^{1:n}$.

Define $\lambda^{j,i} : \{0,1\}^{i-1} \rightarrow \{0,1\}$ as a deterministic function in block $j$ that maps $u^{1:i-1}$ into a bit. Let $\Lambda^{j,i}$  denote the random variable of boolean map $\lambda^{j,i}$ that takes values according to
\begin{equation}
\Lambda^{j,i}(u^{1:i-1})=\left\{
\begin{array}{rcl}
1,       && \text{w.p. }P_{U^{i}|U^{1:i-1}}(1|u^{1:i-1})\\
0,       && \text{w.p. }P_{U^{i}|U^{1:i-1}}(0|u^{1:i-1})
\end{array} \right.
\end{equation}
The maps are chosen prior to the encoding process and are shared by the encoder and the decoders $1$ and $2$.
The coding protocol  is described as follows.
\\\emph{Encoding block $1$:}
\begin{equation}\label{enblock1}
u^i=\left\{
\begin{array}{rcl}
M'_0 \text{ message bits},       && i\in\mathcal{I}_2\\
\lambda^{j,i}(u^{1:i-1}),       && i\in(\mathcal{I}_2)^c
\end{array} \right.
\end{equation}
\emph{Encoding block $j=2,\ldots,k-1$:}
\begin{equation}\label{enblockj}
u^i=\left\{
\begin{array}{rcl}
M'_0 \text{ message bits},       && i\in\mathcal{I}_2\\
\text{message bit in } \mathcal{D}_2, \text{block } j-1,       && i\in\mathcal{D}_{10}\\
M_{11} \text{ message bits},       && i\in\mathcal{D}_{1}\backslash\mathcal{D}_{10}\\
\lambda^{j,i}(u^{1:i-1}),       && i\in(\mathcal{I}_1\cup \mathcal{I}_2)^c\\
\end{array} \right.
\end{equation}
\emph{Encoding block $j=k$:}
\begin{equation}\label{enblockj}
u^i=\left\{
\begin{array}{rcl}
M'_0 \text{ message bits},       && i\in(\mathcal{I}_1\cap\mathcal{I}_2)\\
\text{message bits in } \mathcal{D}_2, \text{block } j-1,       && i\in\mathcal{D}_{10}\\
M_{11} \text{ message bits},       && i\in\mathcal{D}_{1}\backslash\mathcal{D}_{10}\\
\lambda^{j,i}(u^{1:i-1}),       && i\in(\mathcal{I}_1)^c\\
\end{array} \right.
\end{equation}
In each block, the encoder transmits $x^{1:n}=u^{1:n}G^{-1}_n=u^{1:n}G_n$ over the broadcast channel.
Upon receiving the outputs $y_1^{1:n}$ of each block, user $1$ performs successive decoding from block $k$ to block $1$ as follows.
\\\emph{User $1$ decoding block $k$:}
\begin{equation}\label{deblock1user1}
\hat{u}^i=\left\{
\begin{array}{rcl}
\text{arg}\max_{u\in\{0,1\}}P_{U|U^{1:i-1},Y^{1:n}_1}(u|u^{1:i-1},y^{1:n}_1),       && i\in\mathcal{I}_1\\
\lambda^{j,i}(u^{1:i-1}),       && i\in(\mathcal{I}_1)^c
\end{array} \right.
\end{equation}
\emph{User $1$ decoding block $j=k-1,\ldots,2$:}
\begin{equation}\label{deblockjuser1}
\hat{u}^i=\left\{
\begin{array}{rcl}
\text{arg}\max_{u\in\{0,1\}}P_{U|U^{1:i-1},Y^{1:n}_1}(u|u^{1:i-1},y^{1:n}_1),       && i\in\mathcal{I}_1\\
\text{message bits in } \mathcal{D}_{10}, \text{ block } j+1,       && i\in\mathcal{D}_{2}\\
\lambda^{j,i}(u^{1:i-1}),       && i\in(\mathcal{I}_1\cup \mathcal{I}_2)^c\\
\end{array} \right.
\end{equation}
\emph{User $1$ decoding block $j=1$:}
\begin{equation}\label{deblockkuser1}
\hat{u}^i=\left\{
\begin{array}{rcl}
\text{arg}\max_{u\in\{0,1\}}P_{U|U^{1:i-1},Y^{1:n}_1}(u|u^{1:i-1},y^{1:n}_1),       && i\in(\mathcal{I}_1\cap\mathcal{I}_2)\\
\text{message bits in } \mathcal{D}_{10}, \text{ block } j+1,       && i\in\mathcal{D}_{2}\\
\lambda^{j,i}(u^{1:i-1}),       && i\in(\mathcal{I}_2)^c\\
\end{array} \right.
\end{equation}
Upon receiving $y^{1:n}_2$ of each block, user $2$ starts from block $1$ to block $k$.
\\\emph{User $2$ decoding block $j=1$:}
\begin{equation}\label{deblock1user2}
\hat{u}^i=\left\{
\begin{array}{rcl}
\text{arg}\max_{u\in\{0,1\}}P_{U|U^{1:i-1},Y^{1:n}_2}(u|u^{1:i-1},y^{1:n}_2),       && i\in\mathcal{I}_2\\
\lambda^{j,i}(u^{1:i-1}),       && i\in(\mathcal{I}_2)^c
\end{array} \right.
\end{equation}
\emph{User $2$ decoding block $j=2,\ldots,k-1$:}
\begin{equation}\label{deblockjuser2}
\hat{u}^i=\left\{
\begin{array}{rcl}
\text{arg}\max_{u\in\{0,1\}}P_{U|U^{1:i-1},Y^{1:n}_2}(u|u^{1:i-1},y^{1:n}_2),       && i\in\mathcal{I}_2\\
\text{message bits in } \mathcal{D}_2, \text{ block } j-1,       && i\in\mathcal{D}_{10}\\
M_{11} \text{ message bits},       && i\in\mathcal{D}_{1}\backslash\mathcal{D}_{10}\\
\lambda^{j,i}(u^{1:i-1}),       && i\in(\mathcal{I}_1\cup \mathcal{I}_2)^c\\
\end{array} \right.
\end{equation}
\emph{User $2$ decoding block $j=k$:}
\begin{equation}\label{deblockkuser2}
\hat{u}^i=\left\{
\begin{array}{rcl}
\text{arg}\max_{u\in\{0,1\}}P_{U|U^{1:i-1},Y^{1:n}_1}(u|u^{1:i-1},y^{1:n}_1),       && i\in(\mathcal{I}_1\cap\mathcal{I}_2)\\
\text{message bits in } \mathcal{D}_{2}, \text{ block } j-1,       && i\in\mathcal{D}_{10}\\
M_{11} \text{ message bits},       && i\in\mathcal{D}_{1}\backslash\mathcal{D}_{10}\\
\lambda^{j,i}(u^{1:i-1}),       && i\in(\mathcal{I}_1)^c\\
\end{array} \right.
\end{equation}
The average message rates per symbol $(R_0,R_1,R_2)$ in the above coding protocol are given by
\begin{equation}
\begin{split}
R_1+R_0&=R_0'+R_{11}=\frac{1}{kn}[(k-1)|\mathcal{I}_1|+|\mathcal{I}_1\cap \mathcal{I}_2|]\\
&=\frac{(k-1)}{k}I(X;Y_1)+\frac{1}{kn}|\mathcal{I}_1\cap \mathcal{I}_2|+o(1)\\
R_2+R_0&=R_0'=\frac{1}{kn}[(k-1)|\mathcal{I}_2|+|\mathcal{I}_1\cap \mathcal{I}_2|]\\
&=\frac{(k-1)}{k}I(X;Y_2)+\frac{1}{kn}|\mathcal{I}_1\cap \mathcal{I}_2|+o(1).\\
\end{split}
\end{equation}
as $k$ grows, $R_0+R_1$ and $R_0+R_2$ approach arbitrarily closed to $I(X;Y_1)$ and $I(X;Y_2)$ respectively. The decoding complexity $n\log n$ follows from the fact that the likelihood ratio at decoder $m$
\begin{equation}
\begin{split}
L^{i}_{m,n}=\frac{P_{U^i|U^{1:i-1},Y^{1:n}_m}(0|u^{1:i-1},y^{1:n}_m)}{P_{U^i|U^{1:i-1},Y^{1:n}_m}(1|u^{1:i-1},y^{1:n}_m)},~~~~m=1,2.
\end{split}
\end{equation}
can be computed in a recursive manner \cite{arikan2010source}.

The analysis of error probability follows similar steps to those in \cite{honda2013polar,goela2013polar} except that the error probability for user $1$ or $2$ is conditioned on the bits $u^{\mathcal{D}_2}$ or $u^{\mathcal{D}_1}$ respectively known from previous decoded blocks and message side information. The details are omitted here.

\section{BCSI with Common Message and with Noncausal State}
In this section a polar coding scheme is proposed for BCSI with common message and with noncausal state \eqref{channelmodel}. It is also shown that the proposed polar coding scheme achieves the capacity region for degraded BCSI with common message and with noncausal state.

The Gelfand-Pinsker capacity for channel with random state noncausally known at the encoder is given by
\begin{equation}\label{channelwithstate}
C=\max_{p_{U|S}(u|s),x(u,s)}I(U;Y)-I(U;S).
\end{equation}
A straightforward extension of the Gelfand-Pinsker capacity for BCSI with noncausal state is given by \cite{6294445}
\begin{equation}\label{old}
\begin{split}
&R_0+R_1\le I(U;Y_1)-I(U;S),~~~~R_0+R_2\le I(U;Y_2)-I(U;S).
\end{split}
\end{equation}
We now establish an achievable rate region, which is strictly larger than that characterized by \eqref{old}, and present polar codes for achieving the region.
\begin{theorem}
\textup{
For BCSI with common message and with noncausal state \eqref{channelmodel}, where the input has binary alphabet, there exists a polar code sequence with block length $n$ that achieves $(R_0,R_1,R_2)$ if
\begin{equation}\label{betterregion}
\begin{split}
&R_1+R_0\le I(V_1,V_2;Y_1)-I(V_1,V_2;S),\\
&R_2+R_0\le I(V_1;Y_2)-I(V_1;S)
\end{split}
\end{equation}
for binary variables $V_1,V_2$ that satisfy
$(1)$ $(V_1,V_2)\rightarrow (X,S) \rightarrow Y_1$ form a Markov chain,
$(2)$ $(V_1,V_2)\rightarrow (X,S) \rightarrow Y_2$ form a Markov chain,
$(3)$ $I(V_2;Y_1|V_1)>I(V_2;S|V_1)$,
$(4)$ $I(V_1;Y_1)>I(V_1;S)$,
$(5)$ $I(V_1;Y_2)>I(V_1;S)$,
and for some function $f(v_1,v_2,s):~\{0,1\}^2\times \mathcal{S}\rightarrow \mathcal{X}$. As $n$ increases, the encoding and decoding complexity is $O(n\log n)$ and the error probability is $O(2^{-n^\beta})$ for $0<\beta<\frac{1}{2}$.
}
\end{theorem}
\begin{remark}
\textup{The rate region \eqref{betterregion} reduces to \eqref{old} when the random variable $V_2$ remains constant.
}
\end{remark}
\begin{remark}
\textup{Symmetrically, the rate region is achievable if the role of receiver $1$ and receiver $2$ is reversed.
}
\end{remark}
To give an example where the region \eqref{betterregion} is strictly larger than \eqref{old}, consider a broadcast channels with state $(\mathcal{X}\times\mathcal{S},P_{Y_1,Y_2|X,S}(y_1,y_2|x,s),\mathcal{Y}_1\times\mathcal{Y}_2)$ as illustrated in Fig. $3$, with input alphabet $\mathcal{X}=\{1,2,3,4\}$, and state alphabet $\mathcal{S}=\{0,1,2,3,4\}$. Such channel can be viewed as memory with stuck faults with $5$ states. The state $S$ takes values $s=1,2,3,4$ with probability $\frac{p}{4}$ respectively. And $S=0$ with probability $1-p$. The received data $Y_1=S$ when $S=1,2,3,4$. And $Y_1=X$ when $S=0$. The received data $Y_2$ is a blurred version of $Y_1$, where $Y_2=0$ when $Y_1=1,2$, and $Y_2=1$ when $Y_1=3,4$.
\begin{figure}\label{fig:1}
\center
\includegraphics[scale=0.56]{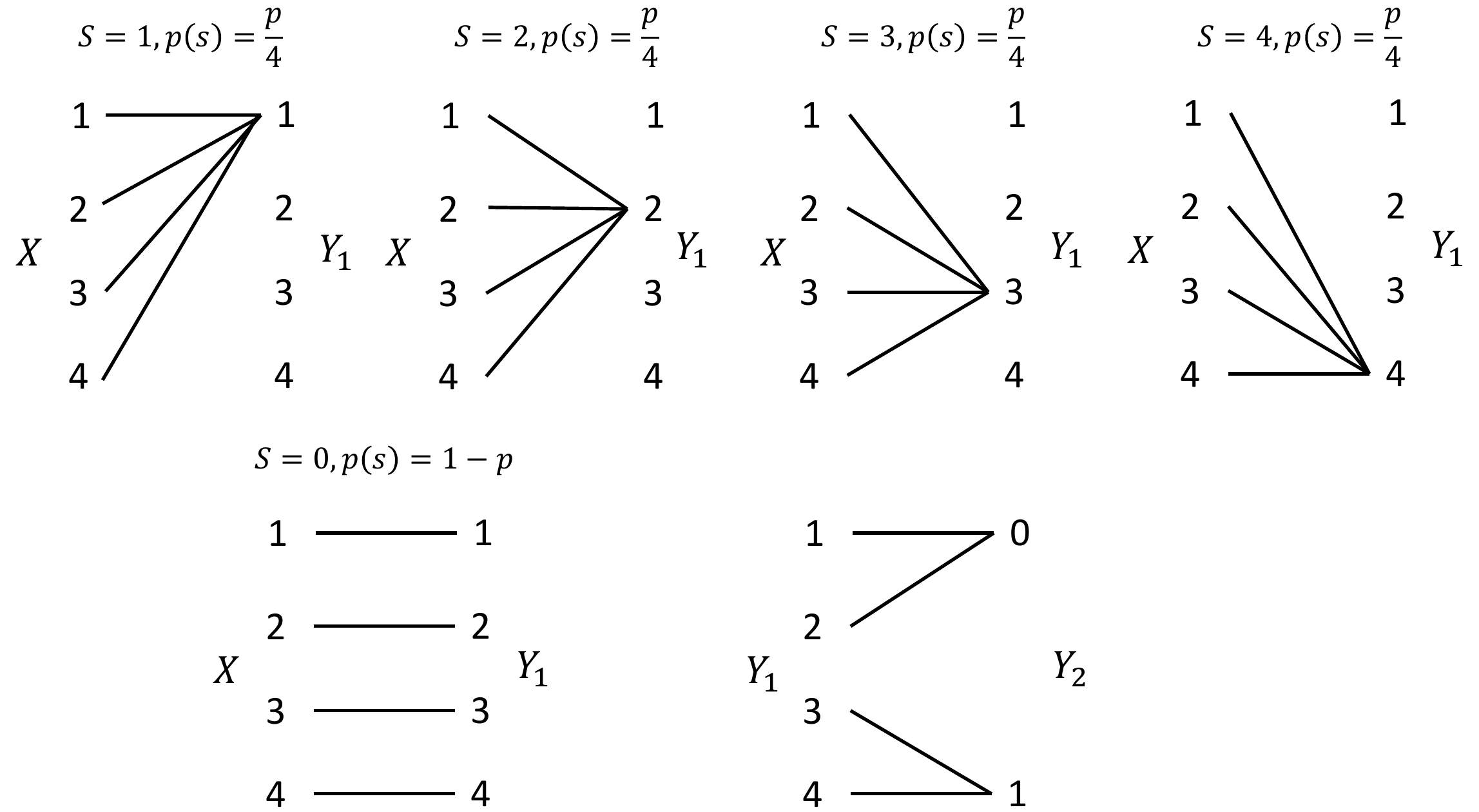} 
\caption{Example of BCSI with noncausal state}
\setlength\belowcaptionskip{0pt}
\end{figure}
\begin{proposition}
\textup{For the broadcast channels with state described above, the rate region \eqref{betterregion} achieves the channel capacity, while the region \eqref{old} is strictly smaller than the channel capacity.}
\end{proposition}
\begin{IEEEproof}
Set the random variable $V_2=S$ when $S=1,2,3,4$, and let $V_2$ be uniformly distributed in $\{1,2,3,4\}$ when $S=0$. Let $V_1$ be a blurred version of $V_2$, where $V_1=0$ if $V_2=1,2$, and $V_1=1$ if $V_2=3,4$. Then set $X=V_2$. It can be verified that the variable $V_1,V_2$ satisfy the conditions $(1)-(5)$ described in Theorem 2. And the rate region \eqref{betterregion} becomes
\begin{equation}\label{crexample}
\begin{split}
&R_1+R_0\le 2-2p,~~~~R_2+R_0\le 1-p.
\end{split}
\end{equation}
It can be proved that the above region \eqref{crexample} is optimal since it achieves the capacities for two separate channels with state where the state is noncausally available at the encoder and the decoder, i.e., $C_1=\max_{p_X(x)}I(X;Y_1|S)=2-2p$ and $C_2=\max_{p_X(x)}I(X;Y_2|S)=1-p$. Furthermore, it can be shown that the region \eqref{old} can not reach the optimal region \eqref{crexample}. Otherwise, if there are random variables $U$ and $X$, such that
\begin{equation}
\begin{split}
&I(U;Y_1)- I(U;S) = H(U|S)-H(U|Y_1)=2-2p,\\
&I(U;Y_2)-I(U;S)= H(U|S)-H(U|Y_2)=1-p.
\end{split}
\end{equation}
Then since $H(U|Y_1)\ge H(U|Y_1,S)$, hence
\begin{equation}\label{equap}
\begin{split}
&H(U|S)-H(U|Y_1,S) \ge H(U|S)-H(U|Y_1)=2-2p,
\end{split}
\end{equation}
On the other hand,
\begin{equation}\label{equap1}
\begin{split}
&H(U|S)-H(U|Y_1,S) = \sum_{s}p(s)[H(U|S=s)-H(U|Y_1,S=s)] \\
&=(1-p)[H(U|S=0)-H(U|Y_1,S=0)]=(1-p)[H(Y_1|S=0)-H(Y_1|U,S=0)]\\
&\le (1-p)H(Y_1)\le 2-2p.
\end{split}
\end{equation}
Hence from \eqref{equap} and \eqref{equap1} we have $H(U|Y_1)=H(U|Y_1,S)$. Similarly, $H(U|Y_2)=H(U|Y_2,S)$. This implies that
\begin{equation}\label{probequa}
\begin{split}
P_{U|Y_2}(u|y_2=0)&=P_{U|Y_2,S}(u|y_2=0,s=1)=P_{U|S}(u|s=1)\\
&=P_{U|Y_2,S}(u|y_2=0,s=2)=P_{U|S}(u|s=2),
\end{split}
\end{equation}
\begin{equation}
\begin{split}
&P_{U|S}(u|s=1)=P_{U|Y_1,S}(u|y_1=1,s=1)=P_{U|Y_1,S}(u|y_1=1,s=0)\\
&=P_{U|S}(u|s=2)=P_{U|Y_1,S}(u|y_1=2,s=2)=P_{U|Y_1,S}(u|y_1=2,s=0).
\end{split}
\end{equation}
According to \eqref{equap} and \eqref{equap1}, $p_{Y_1|S}(y_1|s=0)=\frac{1}{4}$ for $y_1=1,2,3,4$. Therefore, we get
\begin{equation}\label{equation66}
\begin{split}
&P_{Y_1|U,S}(y_1=1|u,s=0)=\frac{P_{U|Y_1,S}(u|y_1=1,s=0)P_{Y_1|S}(y_1=1|s=0)}{P_{U|S}(u|s=0)}\\
&=\frac{P_{U|Y_1,S}(u|y_1=2,s=0)P_{Y_1|S}(y_1=2|s=0)}{P_{U|S}(u|s=0)}=P_{Y_1|U,S}(y_1=2|u,s=0)
\end{split}
\end{equation}
Since $Y_1$ is determined by $(U,S)$, Equation \eqref{equation66} implies that $P_{Y_1|U,S}(y_1=1|u,s=0)=P_{Y_1|U,S}(y_1=2|u,s=0)=0$. Similarly, it can be shown that $P_{Y_1|U,S}(y_1=3|u,s=0)=P_{Y_1|U,S}(y_1=4|u,s=0)=0$, which is a contradiction. Thus the proposition is proved
\end{IEEEproof}
Now we define the sets for polarization and coding. Let $(V_1^{1:n},V_2^{1:n})$ be a sequence of $n$ i.i.d. random variables with pmf $P_{V_1,V_2}(v_1,v_2)$. Set the sequences $U_1^{1:n}=V_1^{1:n}G_n$ and $U_2^{1:n}=V_2^{1:n}G_n$. Define the polarization sets
\begin{equation}
\begin{split}
&\mathcal{H}^{(n)}_{U_1}=\{i\in[n]:Z(U_1^i|U_1^{1:i-1})\ge 1-2^{-n^\beta}\},\\
&\mathcal{L}^{(n)}_{U_1}=\{i\in[n]:Z(U_1^i|U_1^{1:i-1})\le 2^{-n^\beta}\},\\
&\mathcal{H}^{(n)}_{U_1|S}=\{i\in[n]:Z(U_1^i|S^{1:n},U_1^{1:i-1})\ge 1-2^{-n^\beta}\},\\
&\mathcal{L}^{(n)}_{U_1|S}=\{i\in[n]:Z(U_1^i|S^{1:n},U_1^{1:i-1})\le 2^{-n^\beta}\},\\
&\mathcal{H}^{(n)}_{U_1|Y_1}=\{i\in[n]:Z(U_1^i|Y_1^{1:n},U_1^{1:i-1})\ge 1-2^{-n^\beta}\},\\
&\mathcal{L}^{(n)}_{U_1|Y_1}=\{i\in[n]:Z(U_1^i|Y_1^{1:n},U_1^{1:i-1})\le 2^{-n^\beta}\},\\
&\mathcal{H}^{(n)}_{U_1|Y_2}=\{i\in[n]:Z(U_1^i|Y_2^{1:n},U_1^{1:i-1})\ge 1-2^{-n^\beta}\},\\
&\mathcal{L}^{(n)}_{U_1|Y_2}=\{i\in[n]:Z(U_1^i|Y_2^{1:n},U_1^{1:i-1})\le 2^{-n^\beta}\},\\
&\mathcal{H}^{(n)}_{U_2|Y_1,U_1}=\{i\in[n]:Z(U_2^i|Y_1^{1:n},U_1^{1:n},U_2^{1:i-1})\ge 1-2^{-n^\beta}\},\\
&\mathcal{L}^{(n)}_{U_2|Y_1,U_1}=\{i\in[n]:Z(U_2^i|Y_1^{1:n},U_1^{1:n},U_2^{1:i-1})\le 2^{-n^\beta}\}.
\end{split}
\end{equation}
The information sets and the remaining frozen sets for receivers $1$ and $2$ are defined as follows:
\begin{equation}
\begin{split}
&\mathcal{I}_{1}=\mathcal{H}^{(n)}_{U_1|S}\cap\mathcal{L}^{(n)}_{U_1|Y_1},~\mathcal{F}_{1a} = \mathcal{H}^{(n)}_{U_1|S}\cap\{\mathcal{L}^{(n)}_{U_1|Y_1}\}^c,\\
&\mathcal{F}_{1r} = (\mathcal{H}^{(n)}_{U_1|S})^c\cap\{\mathcal{L}^{(n)}_{U_1|Y_1}\}^c,~\mathcal{F}_{1f} = (\mathcal{H}^{(n)}_{U_1|S})^c\cap\{\mathcal{L}^{(n)}_{U_1|Y_1}\},\\
&\mathcal{I}_2=\mathcal{H}^{(n)}_{U_1|S}\cap\mathcal{L}^{(n)}_{U_1|Y_2},~\mathcal{F}_{2a} = \mathcal{H}^{(n)}_{U_1|S}\cap\{\mathcal{L}^{(n)}_{U_1|Y_2}\}^c,\\
&\mathcal{F}_{2r} = (\mathcal{H}^{(n)}_{U_1|S})^c\cap\{\mathcal{L}^{(n)}_{U_1|Y_2}\}^c,~\mathcal{F}_{2f} = (\mathcal{H}^{(n)}_{U_1|S})^c\cap\{\mathcal{L}^{(n)}_{U_1|Y_2}\}.
\end{split}
\end{equation}
\subsection{Polar Codes for the General Gelfand-Pinsker Problem}
Let us now consider
polar codes for realizing the Gelfand-Pinsker binning scheme. Without loss of generality, transmission to receiver $1$ is assumed.
Similar to polar coding for BCSI with common message, block chaining construction is used. Fig. $4$ shows the polar coding scheme, which is briefly stated as follows.
\begin{figure}\label{fig:1}
\center
\includegraphics[scale=0.66]{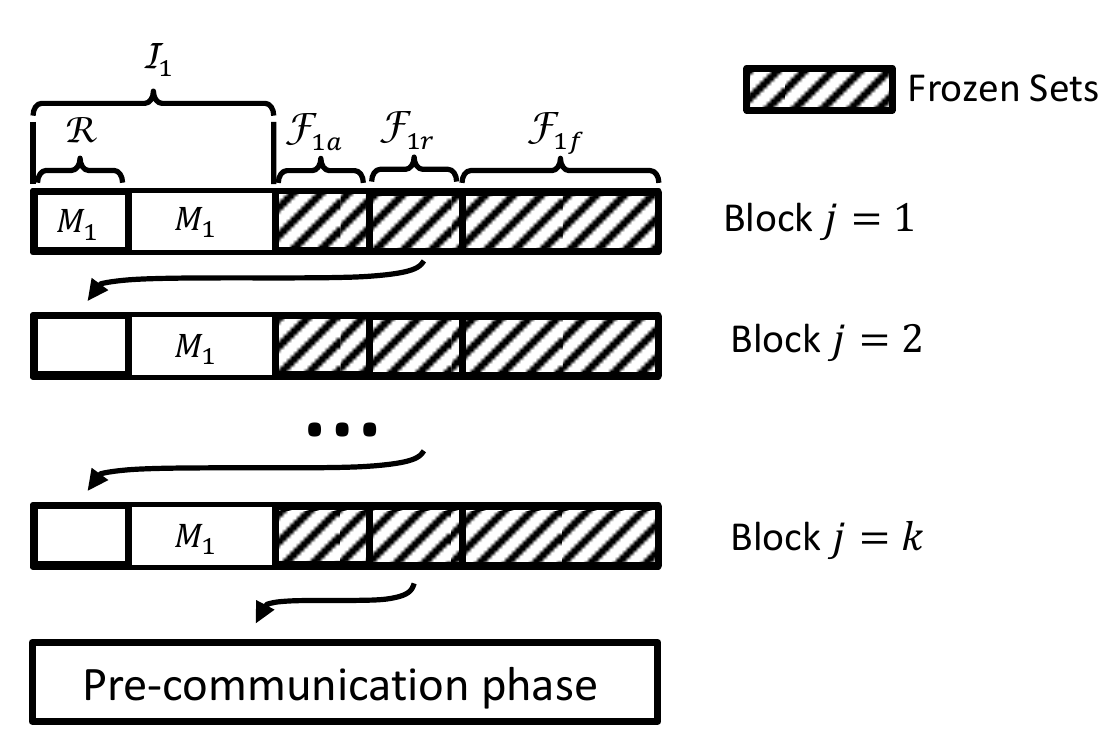} 
\caption{Polar codes for channel with noncausal state.}
\setlength\belowcaptionskip{0pt}
\end{figure}
In block $1$, the encoder puts the message information in the bits $u^{\mathcal{I}_1}$, and generates the remaining frozen bits $u^{\mathcal{I}_1^c}$ using randomly chosen maps with randomness shared between the encoder and the decoders. For block
$j=2,\ldots,k$, the encoder chooses a subset of the information set $\mathcal{R}_1\subseteq\mathcal{I}_1$ and fills the bits $u_1^{\mathcal{R}_1}$ with the information contained in $u^{\mathcal{F}_{1r}}$ of block $j-1$, which is approximately determined by the state sequence $S^n$ and can not be recovered by using the received signal $y_1^{1:n}$. Then the encoder puts information in the bits $u^{\mathcal{I}_1\backslash \mathcal{R}_1}$ and generates the frozen bits $u^{\mathcal{I}_1^c}$ according to randomly chosen maps. Here the bit sets $u^{\mathcal{R}_1}$ in blocks $j=1,\ldots,k$ can be regarded as the chain to transmit the frozen bits $u^{\mathcal{F}_{1r}}$ to user $1$.

Decoder $1$ decodes from block $k$ to block $1$. Note that for block $j=k-1,\ldots,1$, the bits $u_1^{\mathcal{F}_{1r}}$ can be recovered if decoding in block $j+1$ is successful. Since the remaining bits can be recovered either by applying maximum a posteriori rule or by using the randomly chosen maps, decoder $1$ is able to decode the sequence $u^{1:n}$ for block $j=k-1,\ldots,1$ if it decodes $u^{1:n}$ of block $j=k$ successfully. The main difficulty here is the transmission of block $k$.
The work in \cite{gad2014asymmetric} proposed a scheme to transmit the bits of block $k$ by using an extra transmission phase, where state side information is not used at the encoder. There are counterexamples indicating that the scheme in \cite{gad2014asymmetric} may not work. Consider a binary symmetric channel with additive interference $Y=X\oplus Z\oplus S$, where $Z\sim Bern(p)$ and $S\sim Bern(\frac{1}{2})$. It is easy to see that the channel capacity when the encoder does not use the state side information is zero, meaning that the extra phase is not capable of transmitting information. However, when the causal state information is utilized at the encoder, the channel capacity becomes $1-H(p)$, which is nonzero when $0\le p<\frac{1}{2}$. Hence the information can be transmitted.
The following lemma shows that it is sufficient to pre-communicate the bits $u_1^{\mathcal{F}_{1r}}$ of block $k$ by adopting polar coding with causal side information.
\begin{lemma}
\textup{For a channel with random state $(\mathcal{X}\times \mathcal{S}, P_{Y|X,S}(y|x,s), \mathcal{Y})$, where the state is noncausally known at the encoder, if the channel capacity
\begin{equation}
C=\max_{p_{U|S}(u|s),f(u,s)}I(U;Y)-I(U;S)
\end{equation}
is greater than $0$, then
$\max_{p_U(u),f(u,s)}I(U;Y)>0$,
i.e., the capacity for channel with causal state known at the encoder is greater than $0$.
}
\end{lemma}
\begin{IEEEproof}
We first prove that $X\rightarrow S \rightarrow Y$ do not form a Markov chain. Otherwise, we have $p_{Y|S}(y|s)=P_{Y|S,X}(y|s,x)=P_{Y|S,X,U}(y|s,x,u)$ for $P_{S,X,U}(s,x,u)\ne 0$, since $U\rightarrow (S,X) \rightarrow Y$ form a Markov chain. Then $U\rightarrow S \rightarrow Y$ form a Markov chain, which implies that $I(U;S)\ge I(U;Y)$ according to the information processing inequality. This contradicts to the assumption that $C>0$.
Hence, there exist $y_1,~s_1$, and $x_1\ne x_2$,  such that $P_{Y|X,S}(y_1|x_1,s_1)\ne P_{Y|X,S}(y_1|x_2,s_1)$.

For a fixed pmf $P_U(u)$, where $U$ is independent of $S$. choose $u_1 \ne u_2$, such that $P_U(u_1),P_U(u_2) >0$. Let $f(u,s):\mathcal{U}\times \mathcal{S}\rightarrow \mathcal{X} $ be a function such that
\begin{equation}
\begin{split}
&f(u_1,s_1)=x_1,~f(u_2,s_1)=x_2\\
&f(u_1,s)=f(u_2,s)=c\in\mathcal{X},~~~~s\ne s_1
\end{split}
\end{equation}
Setting $x=f(u,s)$, we have
\begin{equation}\label{conditionpr1}
\begin{split}
P_{Y|U}(y_1|u_1)&=\sum_{s,x}P_{S|U}(s|u_1)P_{X|U,S}(x|u_1,s)P_{Y|X,S}(y_1|x,s)\\
&=\sum_{s}P_S(s)P_{Y|X,S}(y_1|f(u_1,s),s)\\
&=\sum_{s\ne s_1}P_S(s)P_{Y|X,S}(y_1|c,s)+P_S(s_1)P_{Y|X,S}(y_1|x_1,s_1).
\end{split}
\end{equation}
Similarly, we have
\begin{equation}\label{conditionpr2}
P_{Y|U}(y_1|u_2)=\sum_{s\ne s_1}P_S(s)P_{Y|X,S}(y_1|c,s)+P_S(s_1)P_{Y|X,S}(y_1|x_2,s_1).
\end{equation}
Now we show that $U$ is not independent of $Y$. Otherwise we have
\begin{equation}
P_{Y|U}(y_1|u_1)=P_{Y}(y_1)=P_{Y|U}(y_1|u_2),
\end{equation}
which is in contradiction with \eqref{conditionpr1} and \eqref{conditionpr2}. Therefore, we conclude that $\max_{p_U(u),f(u,s)}I(U;Y)>0$.
\end{IEEEproof}
To pre-transmit the bits $u_1^{\mathcal{F}_{1r}}$ of block $k$, an extra phase that consists of $t$ blocks is used, where the encoder adopts polar codes for channel with causal state.
The encoder first chooses a random variable $(V',f'(v,s))=\arg {\max _{{P_V}(v),f(v,s)}}I(V;Y)$ and sets the sequence $U'^{1:n}=V'^{1:n}G_n$.
In each block $j=1,\ldots,t$, the bits $u_1^{\mathcal{F}_{1r}}$ of block $k$ are put in locations $\mathcal{I}'_1=\mathcal{H}_{U'}\cap \mathcal{L}_{U'|Y_1}$. And the frozen bits $u^{(\mathcal{I}'_1)^c}$ are generated using randomly chosen maps as usual. Then the encoder transmits $f'(v',s)$ over the channel. Upon decoding, decoder $1$ decodes the sequence $u'^{1:n}$ by applying maximum a posteriori rule and using the randomly chosen maps.
Let $C_{causal}=\max_{P_{V}(v),f(v,s)}I(V;Y)$ be the capacity for channel with state sequence causally available at the encoder. According to Lemma $1$, $C_{causal}> 0$. By fixing $t=\left\lceil \frac{|\mathcal{F}_{1r}|}{C_{causal}}\right\rceil$, the pre-communication of bits $u_1^{\mathcal{F}_{1r}}$ of block $k$ can be completed in $t$ blocks. The average message rate is given by
\begin{equation}
\begin{split}
R_{1}=&\frac{1}{kn+tn}[k(|\mathcal{I}_1|-|\mathcal{R}_1|)+|\mathcal{I}_1\backslash\mathcal{R}_1|]\\
=&\frac{1}{kn+2tn}[k(|\mathcal{H}^{(n)}_{U|S}\cap\mathcal{L}^{(n)}_{U|Y_1}|-|(\mathcal{H}^{(n)}_{U|S})^c\cap (\mathcal{L}^{(n)}_{Y_1|S})^c|)+|\mathcal{I}_1\backslash\mathcal{R}_1|]\\
=&\frac{1}{kn+2tn}[k(|\mathcal{H}^{(n)}_{U}\cap\mathcal{L}^{(n)}_{U|Y_1}\backslash \mathcal{H}^{(n)}_{U}\cap(\mathcal{H}^{(n)}_{U|S})^c|\\
&-|\mathcal{H}^{(n)}_{U}\cap (\mathcal{H}^{(n)}_{U|S})^c\backslash \mathcal{H}^{(n)}_{U}\cap\mathcal{L}^{(n)}_{U|Y_1}|)+|\mathcal{I}_1\backslash\mathcal{R}_1|]\\
=&\frac{1}{kn+2tn}[k(|\mathcal{H}^{(n)}_{U}\cap\mathcal{L}^{(n)}_{U|Y_1}|-|\mathcal{H}^{(n)}_{U}\cap (\mathcal{H}^{(n)}_{U|S})^c|)+|\mathcal{I}_1\backslash\mathcal{R}_1|]\\
=&\frac{k}{k+2t}(I(V;Y_1)-I(V;S))+\frac{1}{kn+2tn}|\mathcal{I}_1\backslash\mathcal{R}_1|+o(1).\\
\end{split}
\end{equation}
As $k$ increases to infinity, the rate $R_1$ approaches $I(V;Y_1)-I(V;S)$. Similar to polar codes for BCSI with common message, the coding complexity is $O(n\log n)$ and the error probability is $O(2^{-n^\beta})$ for any $0<\beta<\frac{1}{2}$.

\subsection{Polar Coding Protocol}
To begin with, split the message $M_1$ into messages $M_{11}$ and $M_{10}$ at rates $R_{11}$ and $R_{10}$ respectively. The coding scheme for BCSI with noncausal state employs a superposition strategy, where the information of $(M_0,M_{10},M_{2})$ is carried by a sequence $u_1^{1:n}$ and the message $M_{11}$ is put in another sequence $u_2^{1:n}$. The encoder transmits $f(v_1,v_2,s)$, where $v_1^{1:n}=u^{1:n}_1G_n$ and $v_2^{1:n}=u^{1:n}_2G_n$. Let the information rates carried by $u_1^{1:n}$ and $u_2^{1:n}$ be given by
\begin{equation}\label{achievablerates}
\begin{split}
R_0+R_{10}&\le I(V_1;Y_1)-I(V_1;S),\\
R_0+R_2&\le I(V_1:Y_2)-I(V_1;S),\\
R_{11}&\le I(V_2;Y_1|V_1)-I(V_2;S|V_1).
\end{split}
\end{equation}
Summing the first and the third inequality in \eqref{achievablerates}, we get \eqref{betterregion}.

Let us first deal with the transmission of the sequence $u_1^{1:n}$, which can be viewed as Gelfand-Pinsker binning simultaneously for the two users. The difficulty here is that the chain construction involves multiple chains. In particular, each decoder $m$ needs a chain to transmit the frozen bits $\mathcal{F}_{mr}$. The two chains must be aligned in a same codeword without conflicts, where a position is assigned with two different values.
To tackle the problem that the two chains may overlap and cause conflicts, we first deal with the case when the two chains do not overlap. Then we show that the case when the two chains overlap can be converted to the first case.

Let us assume that $R_{10}\ge R_2$. The arguments will be similar when $R_{10}\le R_2$. Split the message $M_{10}$ into messages $M_{100}$ and $M_{101}$ at rates $R_{100}$ and $R_{101}$ respectively such that $R_{100}=R_2$. The new equivalent common message is set as
$M_0'=(M_{100}\oplus M_{2},M_0)$. Then we have $R_1+R_0=R_{0}+R_{10}+R_{11}=R_0'+R_{101}+R_{11}$, $R_2+R_0=R_0'$. Set $R'_0=\frac{|\mathcal{I}_2|-|\mathcal{F}_{2r}|}{n}$ and $R'_0+R_{101}=\frac{|\mathcal{I}_1|-|\mathcal{F}_{1r}|}{n}$.
Consider the following two cases: $(a)$ $nR'_0\ge |\mathcal{I}_1\cap \mathcal{I}_2|$.
$(b)$ $nR'_0\le |\mathcal{I}_1\cap \mathcal{I}_2|$.

Case $(a):$ In this case, we can
choose a subset $\mathcal{R}_1\subseteq (\mathcal{I}_1- \mathcal{I}_2)$ and a subset $\mathcal{R}_2\subseteq (\mathcal{I}_2- \mathcal{I}_1)$ such that $|\mathcal{R}_1|=|\mathcal{F}_{1r}|$ and $|\mathcal{R}_2|=|\mathcal{F}_{2r}|$. Similar as in the single user Gelfand-Pinsker case, the subsets $\mathcal{R}_1$ and $\mathcal{R}_2$ act the roles of generating the two chains to transmit the frozen bits $u^{\mathcal{F}_{1r}}$ and $u^{\mathcal{F}_{2r}}$ to the two users respectively.
In case $(a)$ the two chains do not overlap.
Define the sets
\begin{equation}
\begin{split}
\mathcal{M}_{1}&=\mathcal{I}_{1}\backslash\mathcal{R}_{1},~\mathcal{M}_{2}=\mathcal{I}_{2}\backslash\mathcal{R}_{2}\\
\mathcal{D}_{1}&=\mathcal{M}_{1}-\mathcal{M}_2,~\mathcal{D}_2=\mathcal{M}_{2}-\mathcal{M}_{1}.
\end{split}
\end{equation}
Let $\mathcal{D}_{10}\subseteq\mathcal{D}_1$ be a subset of $\mathcal{D}_1$ such that $|\mathcal{D}_{10}|=|\mathcal{D}_2|$.
The coding scheme to transmit $u_1^{1:n}$ is presented in Fig.$5$.
\begin{figure}\label{fig:1}
\center
\includegraphics[scale=0.6]{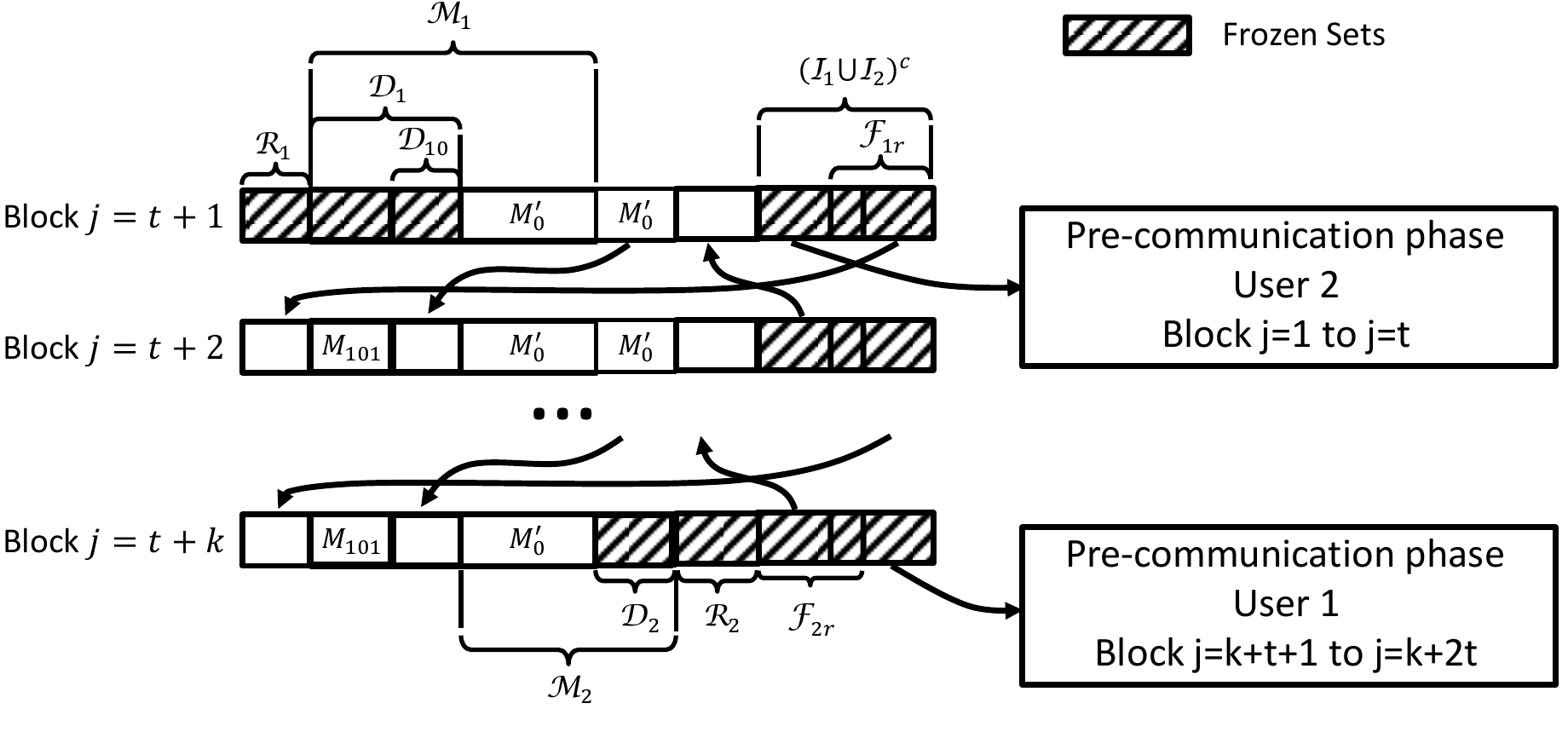} 
\caption{Polar codes for transmitting $u_1^{1:n}$ in case (a).}
\setlength\belowcaptionskip{0pt}
\end{figure}
The first $t$ blocks $j=1,\ldots,t$ are used to pre-communicate the bits $u_1^{\mathcal{F}_{2r}}$ of block $j=t+1$. And the last $t$ blocks $j=k+t+1,\ldots,k+2t$ conveys the bits $u_1^{\mathcal{F}_{1r}}$ of block $j=k+t$.
In block $j=t+1$, the encoder fills the bits $u_1^{\mathcal{R}_2}$ with the information contained in $u_1^{\mathcal{F}_{2r}}$ of block $j+1$ and puts the $M_0'$ information into bits $u_1^{\mathcal{M}_2}$. In block $j=t+2,\ldots,k+t-1$, the encoder copies the bits $u_1^{\mathcal{F}_{2r}}$ of block $j+1$ and the bits $u_1^{\mathcal{F}_{1r}}$ of block $j-1$ to $u_1^{\mathcal{R}_{2}}$ and $u_1^{\mathcal{R}_{1}}$ respectively. The bits $u_1^{\mathcal{D}_{10}}$ are filled with $u_1^{\mathcal{D}_2}$ bits of block $j-1$. The bits $u_1^{\mathcal{D}_1\backslash\mathcal{D}_{10}}$ and bits $u_1^{\mathcal{M}_{2}}$ are inserted with $M_{101}$ bits and $M_{0}'$ bits respectively. In block $j=k+t$, the encoder inserts the positions $\mathcal{R}_1$ with the information contained in $u_1^{\mathcal{F}_{1r}}$ of block $j-1$. The bits $u_1^{\mathcal{D}_{10}}$ are filled with $u_1^{\mathcal{D}_2}$ of block $j-1$ and the bits $u_1^{\mathcal{M}_1\backslash \mathcal{D}_{10}}$ are filled with the information of $M_{101}$. The remaining bits are frozen and generated using randomized maps and the randomness is shared between the encoder and the decoders.

Upon decoding, user $2$ begins by decoding the first $t$ blocks in the pre-communication phase. Then it starts from block $j=t+1$ to block $j=k+t$. For block $t+1$, the bits $u_1^{\mathcal{I}_2\cup\mathcal{F}_{2f}}$ can be decoded by maximum a posteriori rule and the bits $u_1^{\mathcal{F}_{2a}}$ can be recovered using the shared randomized maps. The bits $u_1^{\mathcal{F}_{2r}}$ are pre-communicated through the first $t$ blocks . For block $j=t+2,\ldots,k+t-1$,
The bits $u_1^{\mathcal{F}_{2r}}$, $u_1^{\mathcal{D}_{10}}$, and $u_1^{\mathcal{R}_1}$ can be recovered since the content therein is contained in the bits $u_1^{\mathcal{R}_2}$, $u_1^{\mathcal{D}_2}$, and $u_1^{\mathcal{F}_{1r}}$ respectively decoded in the last block $j-1$. Meanwhile, the bits $u_1^{\mathcal{D}_{1}-\mathcal{D}_{10}}$ is available at user $2$ as side information. The bits $u_1^{\mathcal{I}_2}$ can be decoded based on the received sequence $y^{1:n}_2$. The remaining frozen bits $u_1^{(\mathcal{I}_1\cup\mathcal{I}_2)^c}$ can be calculated using the shared randomized maps. Therefore, user $2$ decodes successfully. In block $j=k$, the decoding of the bits $u_1^{(\mathcal{R}_2)^c}$ is the same as that in block $j=t+2,\ldots,k+t-1$. The bits $u_1^{\mathcal{R}_2}$ are recovered using the randomly chosen maps. Similarly, user $1$ starts from block $k+2t$ to block $t+1$ and is able to decode successfully.

Let $\lambda^{j,i}_{U_1|S} : \{0,1\}^{i-1}\times \mathcal{S}^n \rightarrow \{0,1\}$ be a deterministic map of block $j$. Let $\Lambda^{j,i}_{U_1|S}$ be the random variable of the boolean map $\lambda^{j,i}_{U_1|S}$ that takes values according to
\begin{equation}
\Lambda^{j,i}_{U_1|S}=\left\{
\begin{array}{rcl}
1,       && \text{w.p. }P_{U_1^{i}|U_1^{1:i-1},S^{1:n}}(1|u_1^{1:i-1},s^{1:n})\\
0,       && \text{w.p. }P_{U_1^{i}|U_1^{1:i-1},S^{1:n}}(0|u_1^{1:i-1},s^{1:n})
\end{array} \right.
\end{equation}
Let $\Gamma^j (i)$ be a random variable of function $\gamma^j(i): \{1,\ldots,n\}\rightarrow \{0,1\}$ such that
\begin{equation}
\Gamma^j(i)=\left\{
\begin{array}{rcl}
1,       && \text{w.p. }\frac{1}{2}\\
0,       && \text{w.p. }\frac{1}{2}
\end{array} \right.
\end{equation}
Choose $(V'_1,f'_1(v'_1,s))=\text{arg}\max_{p_{V}(v),f(v,s)}I(V;Y_1)-I(V;S)$ and
$(V'_2,f'_2(v'_2,s))=\text{arg}\max_{p_{V}(v),f(v,s)}$ $I(V;Y_2)-I(V;S)$. Set the sequence $U'^{1:n}_1=V'^{1:n}_1G_n$ and $U'^{1:n}_2=V'^{1:n}_2G_n$.
Let $\Lambda^{j,i}_{U'_m},m=1,2$ be a random variable of function $\lambda^{j,i}_{U'_m}: \{0,1\}^{i-1}\rightarrow \{0,1\}$ such that
\begin{equation}
\Lambda^{i}_{U'_m}=\left\{
\begin{array}{rcl}
1,       && \text{w.p. }P'_{{U'}^{i}_m|U'^{1:i-1}_m}(1|u'^{1:i-1}_m)\\
0,       && \text{w.p. }P_{{U'}^{i}_m|U'^{1:i-1}_m}(0|u'^{1:i-1}_m)
\end{array} \right.
\end{equation}
For chosen functions $\lambda^{j,i}_{U_1|S}$, ${\lambda}^{j,i}_{U'_1}$, and ${\lambda}^{j,i}_{U'_2}$, the encoding procedure is given as follows:
\\\emph{Encoding block $j=1,\ldots,t$:}
\begin{equation}\label{enblockt}
u'^i_2=\left\{
\begin{array}{rcl}
u^{\mathcal{F}_{2r}}_1 \textup{ bits in block } t+1,       && i\in \mathcal{H}_{U'_2} \cap \mathcal{L}_{U'_2|Y_2}\\
{\lambda}^{j,i}_{U'_2}(u'^{1:i-1}_2),       && i\in(\mathcal{H}_{U'_2} \cap \mathcal{L}_{U'_2|Y_2})^c\\
\end{array} \right.
\end{equation}
\emph{Encoding block $j=t+1$:}
\begin{equation}\label{enblockt1}
u^i_1=\left\{
\begin{array}{rcl}
M'_0 \text{ message bits},       && i\in\mathcal{M}_2\\
u^{\mathcal{F}_{2r}}_1 \text{ bits in block } j+1,       && i\in\mathcal{R}_2\\
\lambda^{j,i}_{U_1|S}(u_1^{1:i-1},s^{1:n}),       && i\in(\mathcal{I}_1\cup \mathcal{I}_2)^c\\
\gamma^{j}(i),       && i\in(\mathcal{I}_1 - \mathcal{I}_2)
\end{array} \right.
\end{equation}
\emph{Encoding block $j=k+2,\ldots,k+t-1$:}
\begin{equation}\label{enblockk}
u^i_1=\left\{
\begin{array}{rcl}
M'_0 \text{ message bits},       && i\in\mathcal{M}_2\\
\text{message bits in } \mathcal{D}_2, \text{ block } j-1,       && i\in\mathcal{D}_{10}\\
M_{101} \text{ message bits},       && i\in\mathcal{D}_{1}\backslash\mathcal{D}_{10}\\
u^{\mathcal{F}_{1r}}_1 \textup{ bits in block } j-1,       && i\in\mathcal{R}_{1}\\
u^{\mathcal{F}_{2r}}_1 \textup{ bits in block } j+1,       && i\in\mathcal{R}_{2}\\
\lambda^{j,i}_{U_1|S}(u_1^{1:i-1},s^{1:n}),       && i\in(\mathcal{I}_1\cup \mathcal{I}_2)^c\\
\end{array} \right.
\end{equation}
\emph{Encoding block $j=k+t$:}
\begin{equation}\label{enblockt1}
u^i_1=\left\{
\begin{array}{rcl}
M'_0 \text{ message bits},       && i\in\mathcal{M}_1\cap\mathcal{M}_2\\
\text{message bits in } \mathcal{D}_2, \text{ block } j-1,       && i\in\mathcal{D}_{10}\\
M_{101} \text{ message bits},       && i\in\mathcal{D}_{1}\backslash\mathcal{D}_{10}\\
u^{\mathcal{F}_{2r}}_1 \text{ bits in block } j-1,       && i\in\mathcal{R}_1\\
\lambda^{j,i}_{U_1|S}(u_1^{1:i-1},s^{1:n}),       && i\in(\mathcal{I}_1\cup \mathcal{I}_2)^c\\
\gamma^j(i),       && i\in(\mathcal{I}_2 - \mathcal{I}_1)
\end{array} \right.
\end{equation}
\emph{Encoding block $j=k+t+1,\ldots,k+2t$:}
\begin{equation}\label{enblockt}
u'^i_1=\left\{
\begin{array}{rcl}
u^{\mathcal{F}_{1r}}_1 \textup{ bits in block } k+t,       && i\in \mathcal{H}_{U'_1} \cap \mathcal{L}_{U'_1|Y_1}\\
{\lambda}^{j,i}_{U'_1}(u'^{1:i-1}_1),       && i\in(\mathcal{H}_{U'_1} \cap \mathcal{L}_{U'_1|Y_1})^c\\
\end{array} \right.
\end{equation}
Upon receiving $y^{1:n}_1$ in each block, user $1$ performs successive decoding from block $k+2t$ to block $t+1$ as follows:
\\\emph{User $1$ decoding block $j=k+2t,\ldots,k+t+1$:}
\begin{equation}\label{deblocktuser1}
\hat{u}'^i_1=\left\{
\begin{array}{rcl}
\text{arg}\max_{u'\in\{0,1\}}P_{U'|U'^{1:i-1}_1,Y^{1:n}_1}(u'|u'^{1:i-1}_1,y^{1:n}_1),       && i\in \mathcal{H}_{U'_1} \cap \mathcal{L}_{U'_1|Y_1}\\
{\lambda}^{j,i}_{U'_1}(u'^{1:i-1}_1),       && i\in(\mathcal{H}_{U'_1} \cap \mathcal{L}_{U'_1|Y_1})^c
\end{array} \right.
\end{equation}
\emph{User $1$ decoding block $j=k+t$:}
\begin{equation}\label{deblock1user1}
\hat{u}^i_1=\left\{
\begin{array}{rcl}
\text{arg}\max_{u\in\{0,1\}}P_{U|U^{1:i-1}_1,Y^{1:n}_1}(u|u^{1:i-1}_1,y^{1:n}_1),       && i\in\mathcal{I}_1 \cup \mathcal{F}_{1f}\\
\text{$\hat{u}'^{\mathcal{F}_{1r}}_1$ bits recovered in block } j=k+t+1,\ldots,k+2t,       && i\in\mathcal{F}_{1r}\\
\gamma^j(i),       && i\in  \mathcal{F}_{1a}
\end{array} \right.
\end{equation}
\emph{User $1$ decoding block $j=k+t-1,\ldots,t+2$:}
\begin{equation}\label{deblockjuser1}
\hat{u}^i_1=\left\{
\begin{array}{rcl}
\text{arg}\max_{u\in\{0,1\}}P_{U|U^{1:i-1}_1,Y^{1:n}_1}(u|u^{1:i-1}_1,y^{1:n}_1),       && i\in\mathcal{I}_1\cup \mathcal{F}_{1f}\\
\text{message bits in } \mathcal{R}_{1}, \text{ block } j+1,       && i\in\mathcal{F}_{1r}\\
\text{message bits in } \mathcal{F}_{2r}, \text{ block } j+1,       && i\in\mathcal{R}_{2}\\
\text{message bits in } \mathcal{D}_{10}, \text{ block } j+1,       && i\in\mathcal{D}_{2}\\
\gamma^j(i),       && i\in\mathcal{F}_{1a}- \mathcal{I}_2\\
\end{array} \right.
\end{equation}
\emph{User $1$ decoding block $j=t+1$:}
\begin{equation}\label{deblockjuser1}
\hat{u}^i_1=\left\{
\begin{array}{rcl}
\text{arg}\max_{u\in\{0,1\}}P_{U|U^{1:i-1}_1,Y^{1:n}_1}(u|u^{1:i-1}_1,y^{1:n}_1),       && i\in(\mathcal{I}_1\cap\mathcal{I}_2)\cup\mathcal{F}_{1f}\\
\text{message bits in } \mathcal{R}_{1}, \text{ block } j+1,       && i\in\mathcal{F}_{1r}\\
\text{message bits in } \mathcal{F}_{2r}, \text{ block } j+1,       && i\in\mathcal{R}_{2}\\
\text{message bits in } \mathcal{D}_{10}, \text{ block } j+1,       && i\in\mathcal{D}_{2}\\
\gamma^j(i),       && i\in(\mathcal{I}_1\cup \mathcal{F}_{1a}) - \mathcal{I}_2\\
\end{array} \right.
\end{equation}
Upon receiving $y_2^{1:n}$, decoder $2$ adopts successive decoding in a similar manner as decoder $1$ does. Unlike decoder $1$, decoder $2$ starts from block $1$ to block $k+t$:
\\\emph{User $2$ decoding block $j=1,\ldots,t$:}
\begin{equation}\label{deblocktuser1}
\hat{u}'^i_2=\left\{
\begin{array}{rcl}
\text{arg}\max_{u'\in\{0,1\}}P_{U'|U'^{1:i-1}_2,Y^{1:n}_2}(u'|u'^{1:i-1}_2,y^{1:n}_2),       && i\in\mathcal{H}_{U'_2} \cap \mathcal{L}_{U'_2|Y_2}\\
{\lambda}^{j,i}_{U'_2}(u'^{1:i-1}_2),       && i\in(\mathcal{H}_{U'_2} \cap \mathcal{L}_{U'_2|Y_2})^c
\end{array} \right.
\end{equation}
\emph{User $2$ decoding block $j=t+1$:}
\begin{equation}\label{deblock1user1}
\hat{u}^i_1=\left\{
\begin{array}{rcl}
\text{arg}\max_{u\in\{0,1\}}P_{U|U^{1:i-1}_1,Y^{1:n}_2}(u|u^{1:i-1}_1,y^{1:n}_2),       && i\in\mathcal{I}_2 \cup \mathcal{F}_{2f}\\
\text{$u'^{\mathcal{F}_{2r}}_2$ bits recovered in block } j=1,\ldots,t,       && i\in\mathcal{F}_{2r}\\
\gamma^j(i),       && i\in  \mathcal{F}_{2a}
\end{array} \right.
\end{equation}
\emph{User $2$ decoding block $j=t+2,\ldots,k+t-1$:}
\begin{equation}\label{deblockjuser1}
\hat{u}^i_1=\left\{
\begin{array}{rcl}
\text{arg}\max_{u\in\{0,1\}}P_{U|U^{1:i-1}_1,Y^{1:n}_2}(u|u^{1:i-1}_1,y^{1:n}_2),       && i\in\mathcal{I}_2\cup \mathcal{F}_{2f}\\
\text{message bits in } \mathcal{R}_{2}, \text{ block } j-1,       && i\in\mathcal{F}_{2r}\\
\text{message bits in } \mathcal{F}_{1r}, \text{ block } j-1,       && i\in\mathcal{R}_{1}\\
\text{message bits in } \mathcal{D}_{2}, \text{ block } j-1,       && i\in\mathcal{D}_{10}\\
M_{101} \text{ message bits},       && i\in\mathcal{D}_{1}-\mathcal{D}_{10}\\
\gamma^j(i),       && i\in\mathcal{F}_{2a}- \mathcal{I}_1\\
\end{array} \right.
\end{equation}
\emph{User $2$ decoding block $j=k+t$:}
\begin{equation}\label{deblockjuser1}
\hat{u}^i_1=\left\{
\begin{array}{rcl}
\text{arg}\max_{u\in\{0,1\}}P_{U|U^{1:i-1}_1,Y^{1:n}_2}(u|u^{1:i-1}_1,y^{1:n}_2),       && i\in(\mathcal{I}_1\cap\mathcal{I}_2)\cup\mathcal{F}_{2f}\\
\text{message bits in } \mathcal{R}_{2}, \text{ block } j-1,       && i\in\mathcal{F}_{2r}\\
\text{message bits in } \mathcal{F}_{1r}, \text{ block } j-1,       && i\in\mathcal{R}_{1}\\
\text{message bits in } \mathcal{D}_{2}, \text{ block } j-1,       && i\in\mathcal{D}_{10}\\
M_{101} \text{ message bits},       && i\in\mathcal{D}_{1}-\mathcal{D}_{10}\\
\gamma^j(i),       && i\in(\mathcal{I}_2 \cup \mathcal{F}_{2a})- \mathcal{I}_1\\
\end{array} \right.
\end{equation}

Case $(b):$ In this case,
 $|\mathcal{F}_{2r}|> |\mathcal{I}_2- \mathcal{I}_1|$, which implies that $\mathcal{R}_2\cap\mathcal{I}_1\ne \emptyset$ for any subset $\mathcal{R}_2\in \mathcal{I}_2$ with $|\mathcal{R}_2|=|\mathcal{F}_{2r}|$.
Hence in this case the two chains may overlap with each other.
To avoid the value assignment conflicts in the overlapped set, the main idea is to let the bits $u_1^{\mathcal{R}_2\cap \mathcal{I}_1}$ carry the information contained in $u_1^{\mathcal{R}_2}$ and $u_1^{\mathcal{I}_1}$ simultaneously.
Let $W'_{1}$ and $W'_2$ be a subset of information carried in $(M_{101},u_1^{\mathcal{R}_1})$ and $u_1^{\mathcal{R}_2}$ respectively such that $\log_2|W'_1|=\log_2|W'_2|=|\mathcal{I}_1\cap\mathcal{I}_2|-nR'_0$. Let $M''_0=(M'_0,W'_{1}\oplus W'_{2})$, where $W'_{1}\oplus W'_{2}$ is the bitwise XOR of $W'_1$ and $W'_2$. Since $R''_0=\frac{|\mathcal{I}_1\cap\mathcal{I}_2|}{n}$, we can adopt the coding scheme of case $(a)$, by regarding $M''_0$ as the new equivalent common message. 
Note that in block $j=t+1$, the bits $u_1^{\mathcal{R}_1}$ does not contain information. Hence decoder $2$ can recover $W'_1$ and thus the information contained in $W'_2$.
For blocks $j=t+2,\ldots,k+t$, decoder $2$ knows the information of $(M_{101},u_1^{\mathcal{R}_1})$ since $u_1^{\mathcal{R}_1}$ copies the bits $u_1^{\mathcal{F}_{1r}}$ from block $j-1$. Hence decoder $2$ can recover the information contained in $W'_2$. Similarly, decoder $1$ can recover the information contained in $W'_1$. The message rates $(R_{0},R_{10},R_2)$ are given by
\begin{equation}\label{r10}
\begin{split}
R_0+R_{10}&=\frac{1}{kn+2tn}[(k-1)(|\mathcal{I}_1|-|\mathcal{R}_1|)+|\mathcal{M}_1\cap \mathcal{M}_2|]\\
&=\frac{1}{kn+2tn}[(k-1)(|\mathcal{H}^{(n)}_{U|S}\cap\mathcal{L}^{(n)}_{U|Y_1}|-|(\mathcal{H}^{(n)}_{U|S})^c\cap (\mathcal{L}^{(n)}_{Y_1|S})^c|)+|\mathcal{M}_1\cap \mathcal{M}_2|]\\
&=\frac{k-1}{k+2t}(I(V_1;Y_1)-I(V_1;S))+\frac{1}{k}|\mathcal{M}_1\cap \mathcal{M}_2|+o(1)\\
R_0+R_2&=\frac{1}{kn+2tn}[(k-1)(|\mathcal{I}_2|-|\mathcal{R}_2|)+|\mathcal{M}_1\cap \mathcal{M}_2|]\\
&=\frac{k-1}{k+2t}(I(V_1;Y_2)-I(V_1;S))+\frac{1}{k}|\mathcal{M}_1\cap \mathcal{M}_2|+o(1)
\end{split}
\end{equation}
The transmission of sequence $u_2^{1:n}$ can be regarded as Gelfand-Pinsker binning for user $1$. Define 
\begin{equation}
\begin{split}
&\mathcal{I}_{11}=\mathcal{H}^{(n)}_{U_2|S,U_1}\cap\mathcal{L}^{(n)}_{U_2|Y_1,U_1},~\mathcal{F}_{11a} = \mathcal{H}^{(n)}_{U_2|S,U_1}\cap\{\mathcal{L}^{(n)}_{U_2|Y_1,U_1}\}^c\\
&\mathcal{F}_{11r} = (\mathcal{H}^{(n)}_{U_2|S,U_1})^c\cap\{\mathcal{L}^{(n)}_{U_2|Y_1,U_1}\}^c,~\mathcal{F}_{11f} = (\mathcal{H}^{(n)}_{U_2|S,U_1})^c\cap\{\mathcal{L}^{(n)}_{U_2|Y_1,U_1}\}\\
\end{split}
\end{equation}
Let $\Lambda^{j,i}_{U_2|S,U_1}$ be the random variable of the boolean function
$\lambda^{j,i}_{U_2|S,U_1} : \{0,1\}^{i-1+n}\times\mathcal{S}^n \rightarrow \{0,1\}$ that takes values according to
\begin{equation}
\Lambda^{j,i}_{U_2|S,U_1}=\left\{
\begin{array}{rcl}
1,       && \text{w.p. }P_{U_2^{i}|U_2^{1:i-1},S^{1:n},U_1^{1:n}}(1|u_2^{1:i-1},s^{1:n},u_1^{1:n})\\
0,       && \text{w.p. }P_{U_2^{i}|U_2^{1:i-1},S^{1:n},U_1^{1:n}}(0|u_2^{1:i-1},s^{1:n},u_1^{1:n})
\end{array} \right.
\end{equation}
The encoder uses $t$ blocks as pre-communication phase and
transmits $M_{11}$ through $k$ blocks. Choose a subset $\mathcal{R}_{11}\subseteq\mathcal{I}_{11}$ such that $|\mathcal{R}_{11}|=|\mathcal{F}_{11r}|$.
The coding procedure is given as follows.
\\\emph{Encoding block $j=t+1$:}
\begin{equation}\label{enblockt1}
u^i_2=\left\{
\begin{array}{rcl}
M_{11} \text{ message bits},       && i\in\mathcal{I}_{11}\\
\lambda^{j,i}_{U_2|S,U_1}(u^{1:i-1}_2,u^{1:n}_1,s^{1:n}),       && i\in\mathcal{F}_{11r}\cup \mathcal{F}_{11f}\\
\gamma^{j}(i),       && i\in\mathcal{F}_{11a}\\
\end{array} \right.
\end{equation}
\emph{Encoding block $j=t+2,\ldots,k+t$:}
\begin{equation}\label{enblockt1}
u^i_2=\left\{
\begin{array}{rcl}
M_{11} \text{ message bits},       && i\in\mathcal{M}_{11}\\
u^{\mathcal{F}_{2r}}_2 \text{ bits in block } j-1,       && i\in\mathcal{R}_{11}\\
\lambda^{j,i}_{U_2|S,U_1}(u^{1:i-1}_2,u^{1:n}_1,s^{1:n}),       && i\in\mathcal{F}_{11r}\cup \mathcal{F}_{11f}\\
\gamma^j(i),       && i\in\mathcal{F}_{11a}\\
\end{array} \right.
\end{equation}
\emph{Encoding block $j=k+t+1,\ldots,k+2t$:}
\begin{equation}\label{enblockt}
u'^i_1=\left\{
\begin{array}{rcl}
u^{\mathcal{F}_{11r}}_2 \textup{ bits in block } k+t,       && i\in \mathcal{H}_{U'_1} \cap \mathcal{L}_{U'_1|Y_1}\\
{\lambda}^{j,i}(u'^{1:i-1}_1),       && i\in(\mathcal{H}_{U'_1} \cap \mathcal{L}_{U'_1|Y_1})^c\\
\end{array} \right.
\end{equation}
User $1$ performs successive decoding from block $k+2t$ to block $t+1$ as follows.
\\\emph{User $1$ Decoding block $j=k+2t,\ldots,k+t+1$:}
\begin{equation}\label{deblocktuser1}
\hat{u}'^i_1=\left\{
\begin{array}{rcl}
\text{arg}\max_{u'\in\{0,1\}}P_{U'|U'^{1:i-1}_1,Y^{1:n}_1}(u'|u'^{1:i-1}_1,y^{1:n}_1),       && i\in \mathcal{H}_{U'_1} \cap \mathcal{L}_{U'_1|Y_1}\\
{\lambda}^{j,i}_{U'_1}(u'^{1:i-1}_1),       && i\in(\mathcal{H}_{U'_1} \cap \mathcal{L}_{U'_1|Y_1})^c
\end{array} \right.
\end{equation}
\emph{User $1$ Decoding block $k+t$:}
\begin{equation}\label{deblock1user1}
\hat{u}^i_2=\left\{
\begin{array}{rcl}
\text{arg}\max_{u_2\in\{0,1\}}P_{U^i_2|U^{1:i-1}_2,U^{1:n}_1,Y^{1:n}_1}(u^i_2|u^{1:i-1}_2,u^{1:n}_1,y^{1:n}_1),      && i\in\mathcal{I}_{11} \cup \mathcal{F}_{11f}\\
\text{$\hat{u}'^{\mathcal{F}_{11r}}_1$ bits recovered in block } j=k+t+1,\ldots,k+2t,       && i\in\mathcal{F}_{11r}\\
\gamma^j(i),       && i\in  \mathcal{F}_{11a}
\end{array} \right.
\end{equation}
\emph{User $1$ Decoding block $j=k+t-1,\ldots,t+1$:}
\begin{equation}\label{deblockjuser1}
\hat{u}^i_2=\left\{
\begin{array}{rcl}
\text{arg}\max_{u_2\in\{0,1\}}P_{U^i_2|U^{1:i-1}_2,U^{1:n}_1,Y^{1:n}_1}(u^i_2|u^{1:i-1}_2,u^{1:n}_1,y^{1:n}_1),      &&i\in\mathcal{I}_{11}\cup\mathcal{F}_{11f}\\
\text{message bits in } \mathcal{R}_{11}, \text{ block } j+1,       && i\in\mathcal{F}_{11r}\\
\gamma^j(i),       && i\in\mathcal{F}_{11a}
\end{array} \right.
\end{equation}
The average rate per symbol $R_{11}$ is given by
\begin{equation}\label{r11}
\begin{split}
R_{11}&=\frac{1}{kn+tn}[k(|\mathcal{I}_{11}|-|\mathcal{R}_{11}|)+|\mathcal{I}_{11}\backslash\mathcal{R}_{11}|]\\
&=\frac{1}{kn+tn}[k(|\mathcal{H}^{(n)}_{U_2|U_1,S}\cap\mathcal{L}^{(n)}_{U_2|U_1,Y_1}|-|(\mathcal{H}^{(n)}_{U_2|U_1,S})^c\cap (\mathcal{L}^{(n)}_{Y_1|S})^c|)+|\mathcal{I}_{11}\mathcal{R}_{11}|]\\
&=\frac{k}{kn+tn}[I(V_2;Y_1|V_1)-I(V_2;S|V_1)+|\mathcal{I}_{11}\backslash\mathcal{R}_{11}|+o(1)].
\end{split}
\end{equation}
Let $C_{causal}$ be $C_{causal}=\max\{\max_{P_{V}(v),x(v,s)}I(V;Y_1), \max_{P_{V}(v),x(v,s)}I(V;Y_2)\}$. According to Lemma $1$, $C_{causal}> 0$.
Choose $t=\min \{$ $\left\lceil \frac{|\mathcal{F}_{1r}|}{C_{causal}}\right\rceil,$ $\left\lceil\frac{|\mathcal{F}_{2r}|}{C_{causal}}\right\rceil,\left\lceil\frac{|\mathcal{F}_{11r}|}{C_{causal}}\right\rceil\}$ to be fixed. Then according to \eqref{r10} and \eqref{r11}, $R_1+R_0$ and $R_2+R_0$ approach arbitrarily closed to $I(V_1,V_2;Y_1)-I(V_1,V_2;S)$ and $I(V_1;Y_2)-I(V_1;S)$ respectively, as $k$ grows to infinity.
As $n$ goes to infinity, the encoding and decoding complexity for each user is $O(n\log n)$.
The error probability is upper bounded by $O(2^{-n^\beta})$ for $0<\beta<\frac{1}{2}$.
\subsection{Degraded BCSI with Common Message and with Noncausal State}
Let us now establish the capacity region for degraded BCSI with common message and with noncausal state. A broadcast channels $P_{Y_1,Y_2|X,S}(y_1,y_2|x,s)$ is physically degraded if
\begin{equation}
P_{Y_2|X,S}(y_2|x,s)=P_{Y_2|Y_1}(y_2|y_1)P_{Y_1|X,S}(y_1|x,s)
\end{equation}
for some distribution $P_{Y_1|Y_2}(y_1|y_2)$, i.e., $(X,S)\rightarrow Y_1\rightarrow Y_2$ form a Markov chain. A broadcast channels $P_{Y_1,Y_2|X,S}(y_1,y_2|x,s)$ is stochastically degraded if
\begin{equation}
P_{Y_2|X,S}(y_2|x,s)=\sum_{y_1\in\mathcal{Y}_1}P_{Y_2|Y_1}(y_2|y_1)P_{Y_1|X,S}(y_1|x,s)
\end{equation}
for some distribution $P_{Y_1|Y_2}(y_1|y_2)$. Since the channel capacity depends only on the conditional marginals $P_{Y_1|X,S}(y_1|x,s)$ and $P_{Y_2|X,S}(y_2|x,s)$, the capacity region of a stochastically degraded BC is the same as that of a corresponding physically degraded BC \cite{cover2012elements}. Hence the notion of physically degraded and stochastically degraded are referred to as degraded, and the degradedness is denoted as $P_{Y_1|X,S}(y_1|x,s)\succ P_{Y_2|X,S}(y_2|x,s)$.
\begin{theorem}
\textup{
Let $\mathcal{R}$ be the set of tuples $(R_0,R_1,R_2)$ that satisfy
\begin{equation}\label{capacityregionstate}
\begin{split}
&R_1+R_0\le I(V_1,V_2;Y_1)-I(V_1,V_2;S),\\
&R_2+R_0\le I(V_1;Y_2)-I(V_1;S)
\end{split}
\end{equation}
for some random variables $V_1,V_2$ such that $(1)$ $I(V_2;Y_1|V_1)>I(V_2;S|V_1)$, and $(2)$ $(V_1,V_2)\rightarrow (S,X) \rightarrow Y_1 \rightarrow Y_2$ form a Markov chain, and for some function $\phi :~\mathcal{V}_1\times\mathcal{V}_2\times\mathcal{S}\rightarrow \mathcal{X}$ such that $x=\phi(v_1,v_2,s)$. Then $\mathcal{R}$ is the capacity region of the degraded BCSI with common message and with noncausal state $(\mathcal{X}\times \mathcal{S}, P_{Y_1,Y_2|X,S}(y_1,y_2|x,s), \mathcal{Y}_1 \times \mathcal{Y}_2)$.
}
\end{theorem}
\begin{IEEEproof}
The achievability of region $\mathcal{R}$ is given in Theorem $2$. To prove the converse,
identify random variables $(V^i_{1},V^i_{2})$ as
\begin{equation}
\begin{split}
&V^i_{1}=(M_0,M_1,M_2,S^{i+1:n},Y^{1:i-1}_2),~V^i_{2}=Y^{1:i-1}_1.
\end{split}
\end{equation}
It can be checked that $(V^i_{1},V^i_{2})\rightarrow (S^i,X^i) \rightarrow Y^i_1 \rightarrow Y^i_2$ forms a Markov chain.
According to Fano's inequality,
\begin{equation}
H(M_0,M_2|Y^{1:n}_2,M_1)\le n\epsilon_n.
\end{equation}
Hence, we have
\begin{equation}\label{upper1}
n(R_0+R_2)\le I(M_0,M_2;Y^{1:n}_2|M_1)+n\epsilon_n \le I(M_0,M_1,M_2;Y^{1:n}_2)+n\epsilon_n
\end{equation}
Following the same arguments in \cite{Pinsker1983writing}, It can be shown that
\begin{equation}\label{upper2}
n(R_0+R_1)\le\sum^n_{i=1}\big(I(V^i_1,V^i_2;Y^i_2)-I(V^i_2,V^i;S^i)\big)+n\epsilon_n
\end{equation}
Noticing that $Y_2$ is a degraded version of $Y_1$, it can be similarly proved that
\begin{equation}
n(R_0+R_1)\le\sum^n_{i=1}\big(I(V^i_1,V^i_2;Y^i_2)-I(V^i_2,V^i;S^i)\big)+n\epsilon_n
\end{equation}
According to \eqref{upper1} and \eqref{upper2}, the inequality in \eqref{capacityregionstate} can be proved following the arguments in \cite{steinberg2005coding}.

Next, we show that $I(V_2;Y_1|V_1)>I(V_2;S|V_1)$. Otherwise let $V'_2=\emptyset$, we have
\begin{equation}
\begin{split}
&I(V_1;Y_2)-I(V_1;S)=I(V_1;Y_2)-I(V_1;S),\\
&I(V_1,V'_2;Y_2)-I(V_1,V'_2;S)\ge I(V_1,V_2;Y_2)-I(V_1,V_2;S),
\end{split}
\end{equation}
which yields a larger rate region.
Finally, based on similar arguments that using functional representation lemma \cite{el2011network}, it can be shown that it is sufficient to take $X$ as a deterministic function of $(V_1,V_2,S)$. The theorem is proved.
\end{IEEEproof}
Theorem $3$ can be applied in cellular communication systems. As an example, consider a $4$-user cell as depicted in Fig. $6$, where the base station serves the communication of two pairs of users that wish to exchange information with their partners. Since the users know side information about their own messages, the base station can perform network coding, i.e., pairwise XOR operation of messages, in the downlink transmission so as to increase the transmission rates.
\begin{figure}\label{fig:1}
\center
\includegraphics[scale=0.51]{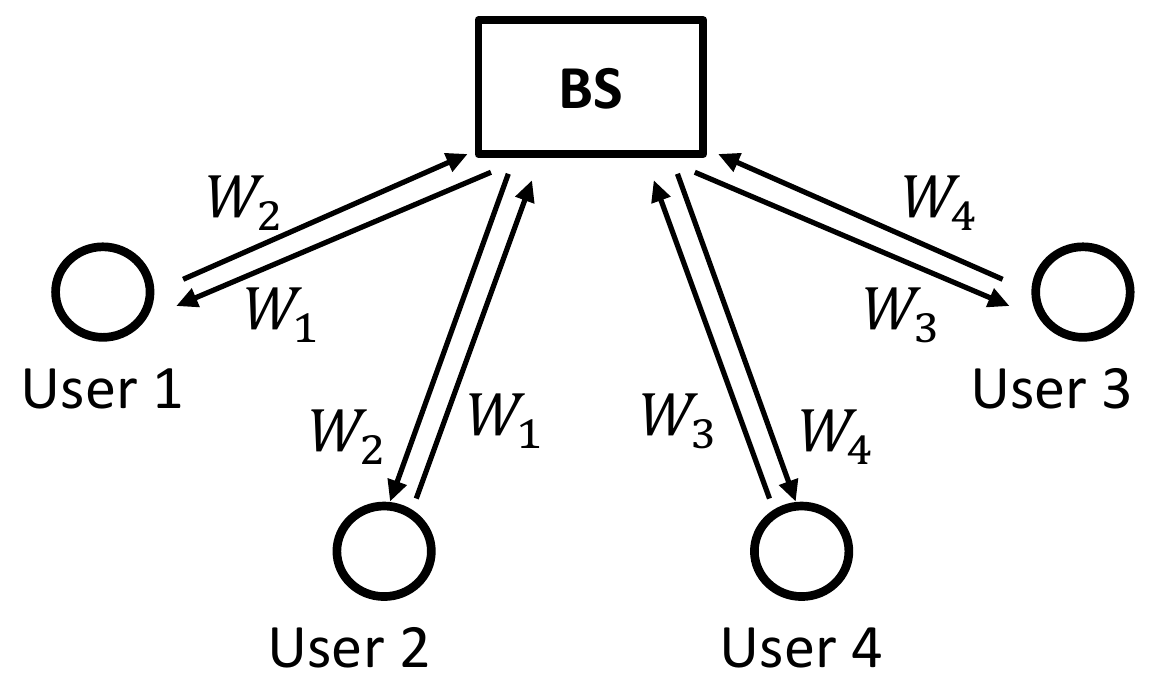} 
\caption{Cellular communication system with two pairwise information exchange tasks.}
\setlength\belowcaptionskip{0pt}
\end{figure}
The downlink transmission is modeled by Gaussian broadcast channels $Y_i=X+Z_i,~i=1,2,3,4$, where $Z_i\sim \mathcal{N}(0,N_i)$ is noise component and the input $X$ has average power $P$.

In superposition coding schemes, the sender may transmit $X=X_1(W_1\oplus W_2)+X_2(W_3\oplus W_4)$, where the pairs of users $(1,2)$ and $(3,4)$ suffer from the interference $X_2(W_3\oplus W_4)$ and $X_1(W_1\oplus W_2)$ respectively. On the other hand, if the base station first generates the signal $X_1(W_1\oplus W_2)$ and then generates $X_2(W_3\oplus W_4)$ by considering $X_1(W_1\oplus W_2)$ as known interference. Then the broadcast channels from base station to users $3$ and $4$ are degraded BCSI with noncausal state. According to Theorem $3$, the base station can achieve the optimal rates for users $3$ and $4$ under interference $X_1(W_1\oplus W_2)$, by choosing proper random variables. Thus the rates for users $3$ and $4$ can be improved compared with superposition coding. The results can also be applied in systems with practical modulation schemes, where $X_1$ and $X_2$ have finite alphabets.

To demonstrate the performance of the proposed scheme, consider a binary symmetric broadcast channels with additive interference $Y_i=X\oplus Z_i\oplus S$, where the interference $S\sim Bern(\frac{1}{2})$ is a Bernoulli random variable and is noncausally available at the encoder. The channel noise $Z_i$ is a Bernoulli random variable $Bern(p_i)$, where it is set $p_1=0.05$, $p_2=0.1$. A polar coding scheme of $k=8$ blocks is assumed.
Fig. 7 plots the error probability of users with respect to the private message rate $R=R_1=R_2$, where the common rate $R_0$ is set to zero.
\begin{figure}\label{fig:1}
\center
\includegraphics[scale=0.18]{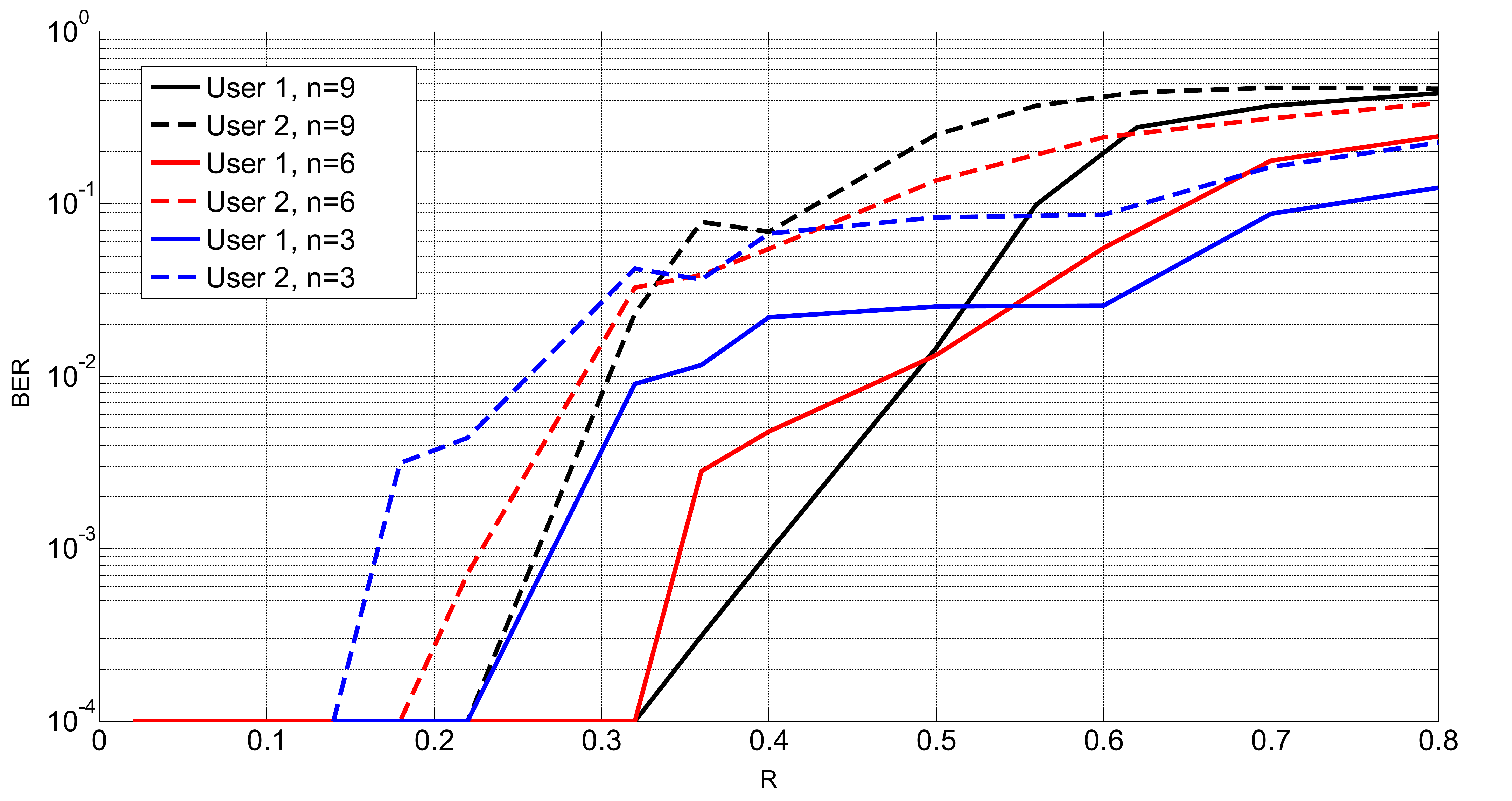} 
\caption{Error probability with respect to message rate $R$.}
\setlength\belowcaptionskip{0pt}
\end{figure}
\section{Conclusion}
In this paper polar coding schemes are proposed for broadcast channels with receiver message side information (BCSI) and with noncausal state available at the encoder. The presented polar coding schemes achieve the performance of encoding/decoding complexity $O(n\log n)$ and error probability $O(2^{-n^\beta})$ for $0<\beta<\frac{1}{2}$. As a special case of the scheme, the capacity for the general Gelfand-Pinsker problem is achieved. It is proved that polar codes are able to achieve the Gelfand-Pinsker capacity through a two-phase transmission. In the first phase the encoder pre-communicates information through polar coding for channel with causal state. In the second phase the encoder transmits messages using chaining construction of polar codes. The presented polar coding scheme for BCSI with common message and with noncausal state has a superposition coding flavor in the sense that the code sequences are successively generated. We use chaining construction to generate the code sequence. In order to let multiple chains share the common information bit indices without conflicts, a nontrivial polarization alignment scheme is proposed. We show that the proposed polar codes achieve the rate region strictly larger than a straightforward extension of the Gelfand-Pinsker result. It is also shown that the presented coding schemes achieve the capacity region for degraded BCSI with common message and with noncausal state.


%

\ifCLASSOPTIONcaptionsoff
  \newpage
\fi




\footnotesize{

\bibliographystyle{IEEEtran}

\bibliography{IEEEabrv,journal}}

\end{document}